\documentclass[sigplan]{acmart}

\acmSubmissionID{2146}
\renewcommand\footnotetextcopyrightpermission[1]{}
\usepackage[most]{tcolorbox}
\usepackage{graphicx}
\usepackage{xcolor}
\usepackage{pifont}
\usepackage{booktabs}
\usepackage{tikz}
\usepackage{subcaption}
\usepackage{url}
\usepackage{algorithm}
\usepackage{algpseudocode}
\usetikzlibrary{patterns}
\usepackage{multirow}
\usepackage{enumitem}

\usepackage{array}
\newcolumntype{C}[1]{>{\centering\arraybackslash}p{#1}}
\usepackage{threeparttable}

\definecolor{symgreen}{RGB}{46, 125, 50}   
\definecolor{symviolet}{RGB}{123, 31, 162} 
\definecolor{symred}{RGB}{198, 40, 40}     

\newcommand{\symP}{
\tikz{\fill[symgreen] (0,0) rectangle (0.15,0.15);}
}
\newcommand{\symS}{
\tikz{
\draw[symviolet, line width=0.4pt] (0,0) rectangle (0.15,0.15);
\fill[symviolet, pattern=north east lines, pattern color=symviolet]
(0,0) rectangle (0.15,0.15);
}
}
\newcommand{\symN}{
\tikz{\draw[symred, line width=0.5pt] (0,0) rectangle (0.18,0.18);}
}

\acmConference[EuroSys'27]{European Conference on Computer
Systems}{April 19--23, 2027}{Rabat, Morocco}

\title[\textsc{SkelOT}: Reusing AOT Compilation Across EVM Contract Families]{\textsc{SkelOT}: Reusing AOT Compilation Across \\ EVM Contract Families}

\thanks{$\triangleright$ Accepted by \textit{European Conference on Computer Systems} (\textcolor{teal}{EuroSys'27}). 
\\$^{\star}$ Most of Z. Wang’s work was made before The University of Manchester.
}

\author{
Sipeng Xie$^1$,
Qianhong Wu$^1$,
Minghang Li$^1$,
Qin Wang$^4$,
Zhipeng Wang$^3$,
Bo Qin$^2$
}
\affiliation{%
\smallskip
  \institution{$^1$\textit{Beihang University} $|$  $^2$\textit{Renmin University of China} \\ $^3$\textit{The University of Manchester} $|$ $^4$\textit{Independent} }
  \country{}
}


\begin{document}

\makeatletter
\fancypagestyle{standardpagestyle}{
  \fancyhf{}
  \fancyfoot[C]{\thepage}
  \renewcommand{\headrulewidth}{0pt}
  \renewcommand{\footrulewidth}{0pt}
}
\fancypagestyle{firstpagestyle}{
  \fancyhf{}
  \fancyfoot[C]{\thepage}
  \renewcommand{\headrulewidth}{0pt}
  \renewcommand{\footrulewidth}{0pt}
}
\pagestyle{standardpagestyle}
\thispagestyle{firstpagestyle}
\makeatother

\begin{abstract}

Ahead-of-time (AOT) compilers (e.g., revmc, evmone, and DTVM) for the Ethereum Virtual Machine (EVM) reuse compilation artifacts at contract-code-hash granularity. This granularity is poorly matched to real EVM workloads dominated by \emph{contract families}: factory-, proxy-, and template-driven deployments that share instruction structure but differ in a small set of embedded constants. Across four EVM chains (Base, Ethereum, BSC, and Arbitrum), we find that 23.1--47.6\% of unique compilable bytecodes map to shared family skeletons within 10K-block windows. Per-hash AOT therefore redundantly recompiles structurally equivalent code, inflating compile time and artifact footprint while reducing workload coverage under finite compile budgets.

We present \textsc{SkelOT}, an AOT framework that lifts the unit of compilation reuse from code hash to family skeleton. \textsc{SkelOT} compiles one native artifact per family, bakes invariant constants into the artifact, and reads variant constants from a per-contract runtime table. Built on revmc/LLVM and evaluated on a 10K-block Base mainnet corpus (3.52M transactions), \textsc{SkelOT} reduces compilation units by 47.5\%, artifact footprint by 57.4\%, and compile time by $2.19\times$, while preserving byte-identical execution outcomes versus per-hash AOT. At runtime, \textsc{SkelOT} delivers a $1.31\times$ median per-contract speedup across family members. Under a compile budget targeting 75\% execution-time coverage, \textsc{SkelOT} needs far fewer artifacts than per-hash AOT, and the advantage holds at every coverage target.

\end{abstract}

\maketitle

\section{Introduction}
\label{sec:intro}

Production ahead-of-time (AOT) compilers for the Ethereum Virtual Machine (EVM) index compilation artifacts by per-contract bytecode hash. Systems such as revmc~\cite{paradigm2024revmc}, evmone~\cite{ipsilon2024evmone}, and DTVM~\cite{zhou2025dtvm} all use this granularity. Each deployed contract is an independent compilation target. Two contracts that differ only in their embedded constants are compiled and cached as separate artifacts. Per-hash AOT, therefore, does not share compiled code across contracts.

Modern EVM workloads, however, are not flat collections of independent contracts. They are dominated by \emph{contract families}: sets of contracts that share instruction structure and differ only in a small set of embedded constants. Factory-driven mass deployment, proxy architectures, and template-based reuse have been documented across the ecosystem~\cite{wang2026factories,sun2023composition,he2020clones}. Naive per-hash AOT compiles each family member separately. The cost is redundant compile work, a native-artifact cache that scales with deployment count rather than with the number of families, and reduced workload coverage under a finite compile budget.

Prior smart-contract research has studied clone prevalence and code reuse~\cite{he2020clones,khan2022clones,sun2023composition}, factory and proxy deployment patterns~\cite{wang2026factories,zhang2025proxy}, and contract similarity detection~\cite{liu2019birthmarks,liu2018eclone}. These works treat contracts as deployment or similarity objects, not as units of compilation work. The compilation-granularity problem has not been raised.

\begin{figure}[H]
    \centering
    \includegraphics[width=0.95\linewidth]{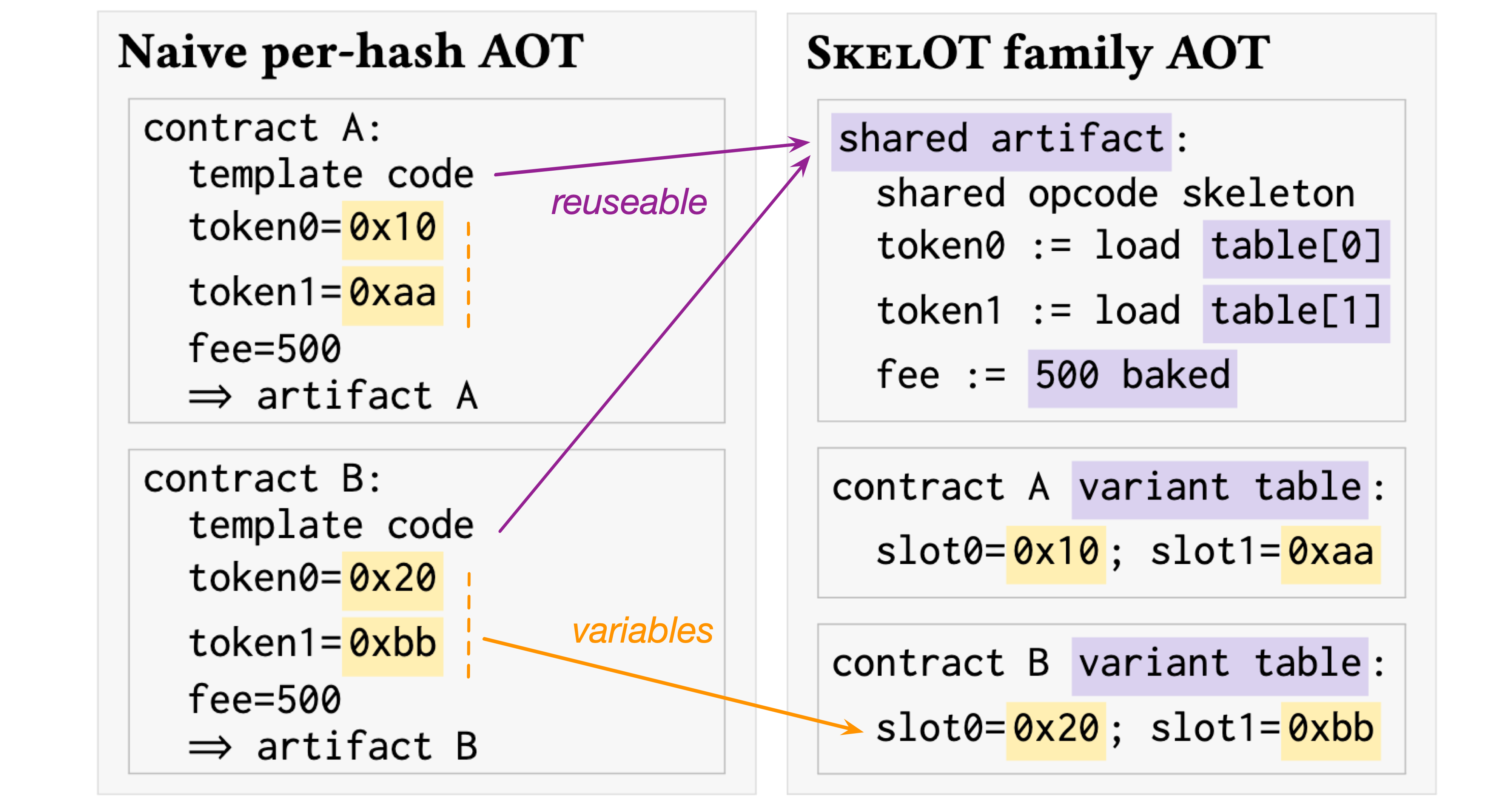}
    \caption{\textbf{\textsc{SkelOT} overview}: Naive AOT (\textit{left}) compiles per code hash; our \textsc{SkelOT} (\textit{right}) compiles once per family and stores only differing constants in per-contract tables.}
    \label{fig:skelot-Intro}
\end{figure}

\vspace{3pt}
In this paper, we reframe the problem. \emph{Exact code hash is a good identity key for a deployed contract but a poor reuse unit for AOT compilation}. The right reuse unit is the \emph{contract-family skeleton}. We define it by what an AOT compiler can reuse across family members, not by what a clone detector identifies as similar.

We present \textsc{SkelOT}, a skeleton-aware AOT framework. \textsc{SkelOT} extracts the shared skeleton from each contract family and applies an \emph{invariant/variant constant split}. It makes this split once per family, by comparing the bytecode of every member. Invariants hold the same value across every family member and are baked into the compiled skeleton as compile-time constants. Variants differ across members and are read at runtime from a per-contract table. One compile therefore covers every member of the family. Under a finite compile budget, this either saves compile time or covers more of the workload.

We implement \textsc{SkelOT} on revmc~\cite{paradigm2024revmc} and evaluate it on four representative EVM-compatible chains, one 10K-block window per chain. Across compile time, native-artifact footprint, runtime dispatch, and budgeted coverage, \textsc{SkelOT} improves over Naive AOT while preserving byte-identical execution outcomes.

Our contributions are summarized as follows:
\begin{itemize}[leftmargin=*]
  \item 
  We identify \textbf{a structural mismatch} between Naive AOT and modern EVM workloads: per-hash compilation follows bytecode identity, whereas family-driven deployment creates many contracts with the same instruction skeleton but different embedded constants. Across four public EVM-compatible networks (Base, Ethereum, BSC, and Arbitrum), 23.1--47.6\% of unique compilable bytecodes collapse into shared family skeletons. We frame the issue as a \textit{compilation-granularity} problem (\S\ref{sec:background}, \S\ref{sec:families}).
  
  \item  We propose \textbf{skeleton extraction} with an invariant/variant constant split as a reuse abstraction. It preserves compile-time optimization for invariant constants while externalizing only the variant subset that differs across family members (\S\ref{sec:design}).
  
  \item We deliver \textbf{an end-to-end implementation} covering skeleton extraction, variance analysis, family-level compilation, runtime constant-table execution, and family-aware caching and dispatch (\S\ref{sec:implementation}).
  
  \item We conduct a large-scale \textbf{empirical validation} on a 10K-block Base mainnet corpus with 3.52M transactions. \textsc{SkelOT} reduces compilation units by 47.5\%, native-artifact footprint by 57.4\%, and compile time by $2.19\times$, while preserving byte-identical outcomes against per-hash AOT across 9{,}997 non-empty blocks. It also achieves a $1.31\times$ median runtime speedup. \textsc{SkelOT} also needs a smaller compile budget to cover a given share of execution time than per-hash AOT, with the advantage holding at every coverage target (\S\ref{sec:results}).
\end{itemize}

\section{Why Is Native AOT Insufficient?}
\label{sec:background}

Existing EVM AOT engines compile at code-hash granularity, so deployment identity also becomes the reuse boundary. This boundary is simple but too fine-grained for modern workloads. We show why in this section.

\subsection{EVM Execution And AOT Compilation}
\label{sec:bg-evm}

The chain keeps a global state that holds account balances, contract storage, and the bytecode of every deployed contract. The state advances one block at a time, as each block's transactions execute the bytecode of the contracts they call, reading and writing the rest of the state. Contracts also call other contracts, so a single transaction can expand into a stack of nested calls. Ethereum Virtual Machine~\cite{wood2014ethereum,xie2026mhot} is a stack machine over 256-bit words. A contract's bytecode is a sequence of opcodes. A small subset of those opcodes, the \emph{PUSH} opcodes \texttt{PUSH1} through \texttt{PUSH32}, carry an inline immediate operand of 1 to 32 bytes loaded onto the stack at execution time. Every other opcode acts on values already on the stack. Control flow uses \texttt{JUMP} and \texttt{JUMPI}, whose target must coincide with a \texttt{JUMPDEST} opcode in the same contract, and targets that fail this check trap. A few opcodes also read a contract's bytecode as data, such as \texttt{CODESIZE} and \texttt{CODECOPY}. Each deployed contract is identified on chain by a code hash, the keccak digest of its bytecode. Solidity also appends a metadata trailer to the bytecode, a short fingerprint of the source and compiler settings. The trailer is data and never executes, yet the code hash covers it like every other byte.

Existing EVM AOT engines lower each contract's bytecode into native code without changing this execution model. The host compiler we build on (revmc, sitting on LLVM~\cite{lattner2004llvm}) translates the opcode stream into LLVM IR. It materializes the 256-bit stack as IR-level memory that LLVM can promote to SSA values~\cite{cytron1991ssa}, and lowers arithmetic and stack manipulation directly into IR operations the optimizer can fold and specialize. Memory and storage opcodes, by contrast, become opaque calls to host runtime routines, so the IR pipeline treats them as black-box functions and does not optimize across them. Control flow is reconstructed from the bytecode and lowered into native branches, with one analysis pass folding constant-target jumps into direct branches before the IR is handed to LLVM.

A developer compiles source to bytecode once with solc~\cite{solidity2026docs} before contract deployment. Every node that wants native execution must compile deployed bytecode, and it keeps compiling as new contracts arrive, on the same machine resources that serve execution. Fees are charged by the opcodes of the original bytecode, so native code only changes how fast an opcode runs.

\subsection{Naive Per-Hash AOT}
\label{sec:bg-naive}

We call the conventional design of existing EVM AOT compilers \emph{Naive per-hash AOT}. This baseline compiles each distinct code hash separately. In other words, if two deployed contracts do not have exactly the same bytecode, they are treated as different compilation targets, even when most of their code structure is shared.

This design is simple and safe. A code hash is a clear on-chain identifier, so it gives the compiler an easy cache key and a clean correctness boundary. Each native artifact maps to one deployed contract, and the compiler only needs to iterate over the unique hashes in the workload. The limitation is that code hash is used not only as a deployment identity, but also as the compilation-reuse boundary. As a result, existing AOT systems can reuse native code only when two contracts are byte-for-byte identical, missing reuse opportunities among contracts that share the same structure but differ only in embedded constants.

\subsection{Why Identity-Based Reuse Fails In Practice}
\label{sec:bg-workload}

The assumption fails in practice because three forces shape modern EVM deployment and decouple structural novelty from deployment count. First, contracts are typically built from shared libraries, audited base implementations, and source-level templates, so composition-level reuse propagates almost unchanged into bytecode-level near-duplicates~\cite{sun2023composition,he2020clones}. Second, factory contracts mass-instantiate near-identical logic from a single source, and members of the resulting family differ only in constructor-baked constants such as owner addresses, token identifiers, pool parameters, or fee tiers~\cite{wang2026factories}. Third, proxy-heavy architectures separate deployment identity from implementation, so a single implementation serves many proxy instances, and both the proxy layer and the implementation layer produce clusters of structurally identical contracts in the on-chain workload~\cite{wang2026factories,zhang2025proxy}. Each force produces families of contracts that share instruction structure and differ only in a bounded set of embedded constants. Every member of such a family still carries a distinct code hash, so Naive AOT compiles each member separately.

\subsection{The Threefold Cost Of The Mismatch}
\label{sec:bg-cost}

The cost of the identity-equals-reuse assumption is threefold. 

First, the compiler performs redundant translation and optimization work across near-duplicate family members. It lowers, optimizes, and register-allocates the same instruction sequence once per member rather than once per family. Dynamic-compilation studies have flagged this redundancy as a dominant source of wasted warmup and end-to-end latency across short-lived deployments~\cite{pecimuth2024reusability}. 

Second, the native code cache stores structurally duplicated artifacts. Its footprint grows with the deployment count of each family rather than with the number of distinct families in the workload, so memory and disk store the same compiled logic repeatedly. Production code-cache designs, by contrast, key on identity so that a cache hit substitutes a stored artifact for a full recompile~\cite{wasmtime_cache_docs,v8_codecache_devs}. 

Third, and most consequential in production, redundant compile work reduces how much of the workload the AOT pipeline can cover under a finite budget. Every real deployment compiles under such a budget, whether a wall-clock window before a block is due, a memory ceiling on the compile cache, or a background-compilation cap shared with execution. The JIT-policy literature shows that such budgets force selective compilation, threshold tuning, and explicit tradeoffs between how aggressively methods are compiled and how quickly the hot working set is covered~\cite{jantz2013jitpolicy}. Compile budgets make the first two costs operationally visible. Wasted translation work and duplicated artifacts become uncovered contracts, which fall back to the interpreter and dominate tail latency.

\section{Contract Families in EVM Workloads}
\label{sec:families}

We use \emph{family} to denote a compiler-relevant reuse boundary, formally defined in \S\ref{sec:families:defn}. This section shows that the boundary is not only common in the studied EVM workloads, but also appears where compilation cost is high and remains stable over time.

\subsection{Defining A Contract Family}
\label{sec:families:defn}

Our contract-family definition is driven by compilation reuse. Given an EVM bytecode sequence, its \emph{skeleton} removes the immediate-value bytes of each \texttt{PUSH} instruction, strips Solidity metadata and trailing zero bytes, and keeps all opcodes in their original positions. A \emph{family} is then the set of contracts whose skeleton byte sequences are bytewise identical~\cite{diangelo2024evolution}. \S\ref{sec:design:skeleton} explains why this byte-level representation is suitable for compilation. Family membership is therefore a deterministic byte-equality check, rather than a similarity score.

\begin{itemize}[leftmargin=*]
    \item For example, \texttt{PUSH20 0xaaa... PUSH1 0x05 ADD} and
\texttt{PUSH20 0xbbb... PUSH1 0x05 ADD} share the skeleton
\texttt{PUSH20 * PUSH1 * ADD}; replacing \texttt{ADD} with another opcode would
place the contract in a different family.
\end{itemize}

This bytewise boundary is necessary because \textsc{SkelOT} reuses compiled code, not merely analysis results. Ethereum clone-detection techniques usually cluster contracts under a similarity threshold~\cite{wang2025clonedetection}. Such clusters are useful for software analysis, but they do not identify a safe compiler reuse boundary. The compiler must share one instruction stream while specializing only contract-specific constants. Skeleton equality provides this boundary. Contracts in the same family have the same opcode layout and control-flow structure, and may differ only in immediate values that \textsc{SkelOT} later classifies as invariant or variant (\S\ref{sec:design:split}). Contracts that are only similar, but not skeleton-identical, therefore fall outside \textsc{SkelOT}'s reuse model.

\subsection{Prevalence On The Corpus}
\label{sec:families:prevalence}

We apply the same filter (\S\ref{sec:eval-setup}) to one 10K-block window on each of four EVM-compatible networks: Base~\cite{coinbase2023base}, Ethereum~\cite{wood2014ethereum}, BSC (BNB Smart Chain)~\cite{bnbchain2024overview,li2025does}, and Arbitrum~\cite{offchainlabs2024arbitrum}.

\begin{table}[!]
  \centering
  \scriptsize
  \setlength{\tabcolsep}{3pt}
  \begin{threeparttable}
  \caption{Cross-chain skeleton deduplication.}
  \label{tab:table1_dedup_xchain}
  \begin{tabular}{lcccc}
    \toprule
    \multicolumn{1}{c}{\textbf{Chain} (blocks)} & \#\textbf{codes} & \#\textbf{skels} & \textbf{Redu.} (\%) & \textbf{Window range} (\%) \\
    \midrule
    BASE {\tiny(38{,}004{,}930--38{,}014{,}929)} & 19{,}130 & 10{,}025 & 47.60 & [40.91, 42.42] \\
    ETH \ {\tiny(23{,}000{,}000--23{,}009{,}999)} & 35{,}215 & 24{,}406 & 30.69 & [26.83, 28.43] \\
    BSC \ {\tiny(95{,}897{,}884--95{,}907{,}883)} & 14{,}444 & 9{,}996 & 30.79 & [24.67, 27.13] \\
    ARB {\tiny(458{,}863{,}320--458{,}873{,}319)} & 2{,}895 & 2{,}225 & 23.14 & [16.34, 21.03] \\
    \bottomrule
  \end{tabular}
  \begin{tablenotes}[flushleft]
    \scriptsize
    \item[-]  Reduction (Redu.) is computed over each 10K-block corpus. Window range reports ten within-chain 1K-block sub-windows.
  \end{tablenotes}
  \end{threeparttable}
\end{table}

Table~\ref{tab:table1_dedup_xchain} lists the inspected block ranges and the per-chain corpus and dedup statistics. On every chain, the corpus of unique compilable bytecodes collapses onto a substantially smaller set of skeletons, yielding 10K-window reductions of 
47.60\% on Base, 30.69\% on Ethereum, 30.79\% on BSC, and 23.14\% on Arbitrum. The reduction also stays bounded across each chain's $n{=}10$ non-overlapping 1K-block sub-windows, with ranges of $[40.91\%, 42.42\%]$ on Base, $[26.83\%, 28.43\%]$ on Ethereum, $[24.67\%, 27.13\%]$ on BSC, and $[16.34\%, 21.03\%]$ on Arbitrum. These bounded ranges show that the reduction is uniform across sub-windows rather than driven by one deployment burst. The corpus-level rate sits above every sub-window because a larger window merges more members into each family. Family structure is therefore a property of the EVM workload, not of any one chain.

\subsection{Where Family Structure Concentrates}
\label{sec:families:variance}

\begin{table}[t]
  \centering
  \scriptsize
  \begin{threeparttable}
  \caption{Size-stratified skeleton deduplication on the 10K-block Base corpus.}
  \label{tab:table2_dedup}
  \vspace{-0.6\baselineskip}
  \begin{tabular}{ccccc}
    \toprule
   \textbf{Bucket} & \#\textbf{bytecode} & \#\textbf{skeletons} & \textbf{Reduction} (\%) & \textbf{Byte share} (\%) \\
    \cmidrule(lr){1-5}
    tiny   &    570 &   166 & 70.9 ($\pm$10.1) & 0.01 \\
    small  &  1{,}204 &   513 & 57.4 ($\pm$8.0) & 0.27 \\
    medium &  4{,}052 & 2{,}609 & 35.6 ($\pm$5.6) & 5.21 \\
    large  &  6{,}114 & 4{,}256 & 30.4 ($\pm$3.5) & 26.09 \\
    huge   &  7{,}190 & 2{,}484 & 65.5 ($\pm$5.8) & 68.41 \\
    \cmidrule(lr){2-5}
    \textbf{Total}         & \textbf{19{,}130} & \textbf{10{,}025} & \textbf{47.6} & \textbf{100.00} \\
    \bottomrule
  \end{tabular}
  \begin{tablenotes}[flushleft]
    \scriptsize
    \item[-] Reduction is reported per bucket with population std.\ dev.\ over ten 1K-block windows. Byte share is the share of raw bytecode bytes.
    \item[-] Buckets follow member bytecode length, so a skeleton whose members straddle a boundary counts once in each bucket it spans; the skeleton rows sum to 10{,}028, three more than the 10{,}025 distinct skeletons.
  \end{tablenotes}
  \vspace{-0.8\baselineskip}
  \end{threeparttable}
\end{table}

A corpus-wide reduction would be less relevant to compilation if it were driven mainly by tiny proxies or stubs. On the Base corpus, it is not. We bucket bytecode by length into tiny ($<$100~B), small (100--999~B), medium (1,000--4,999~B), large (5,000--14,999~B), and huge ($\geq$15,000~B) contracts. Skeleton reduction is strongest in the largest contracts, where any eventual compile-side reuse is most valuable. The huge bucket accounts for 68.41\% of raw input bytes and reduces by 65.5\%, while the tiny bucket accounts for only 0.01\% of raw input bytes. Table~\ref{tab:table2_dedup} reports the full per-bucket breakdown with across-window variability and input-byte share. \S\ref{sec:results} then quantifies compile-side savings on this corpus.

\subsection{Family-Size Distribution}
\label{sec:families:distribution}

Across Base, Ethereum, BSC, and Arbitrum, a small number of large families produce the reduction, not many accidental near-duplicates. The top families correspond to recognizable, long-lived protocol and deployment templates. They include Uniswap~V3~\cite{adams2021uniswap} and PancakeSwap V3~\cite{pancakeswap2023contracts} pool skeletons, PancakeSwap V3 liquidity-mining pool variants~\cite{pancakeswap2023contracts}, EIP-1167~\cite{murray2018eip1167} minimal proxies, TransparentUpgradeableProxy families~\cite{openzeppelin2024proxy}, ClankerToken ERC-20~\cite{vogelsteller2015erc20,clanker2024clankertoken} deployments, and Ethereum ERC1155Creator proxy/NFT-collection families~\cite{radomski2018eip1155,manifold2024creator}. These templates appear repeatedly across the top-family lists and recur in the deployment stream rather than appearing as isolated duplicates.

\begin{figure}[tbp]
  \centering
  \includegraphics[width=\columnwidth]{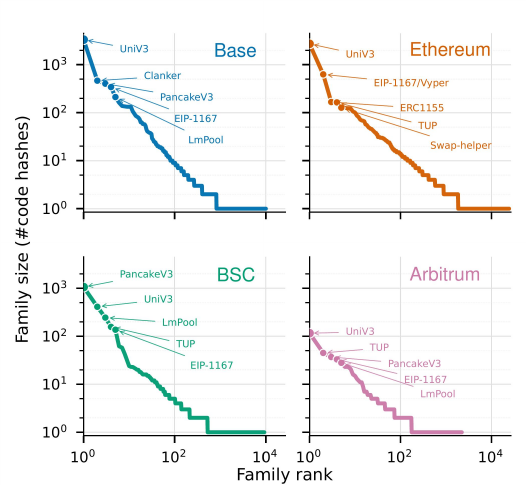}
  \caption{Four-chain family-size distribution}
  \label{fig:family_distribution_xchain}
\end{figure}

Figure~\ref{fig:family_distribution_xchain} shows the four chains as matched log--log rank-size panels; each panel uses arrow callouts for the top families on that chain, and the heavy-tailed shape recurs on every corpus. At the exact-code level, 1{,}672 code hashes recur on at least two chains and 145 recur on all four; at the skeleton level, the counts rise to 2{,}970 and 302. Thus the recurring pattern is not only byte-for-byte redeployment. It is shared contract structure across chains. Every cross-chain code hash maps to the same skeleton, which serves as a consistency check on the extraction.

The singleton tail marks the boundary of family-level reuse. For instance, on the Base corpus, 9{,}200 of the 10{,}025 distinct skeletons appear only once. These singleton skeletons fall back to Naive AOT because there is no cross-member disagreement from which to derive variant positions. The reuse opportunity comes from the 825 multi-member families, which cover 9{,}930 unique compilable bytecodes, or 51.9\% of the corpus.

\section{SkelOT Design}
\label{sec:design}

\begin{figure}[!t]
  \centering
  \includegraphics[
    width=\columnwidth]{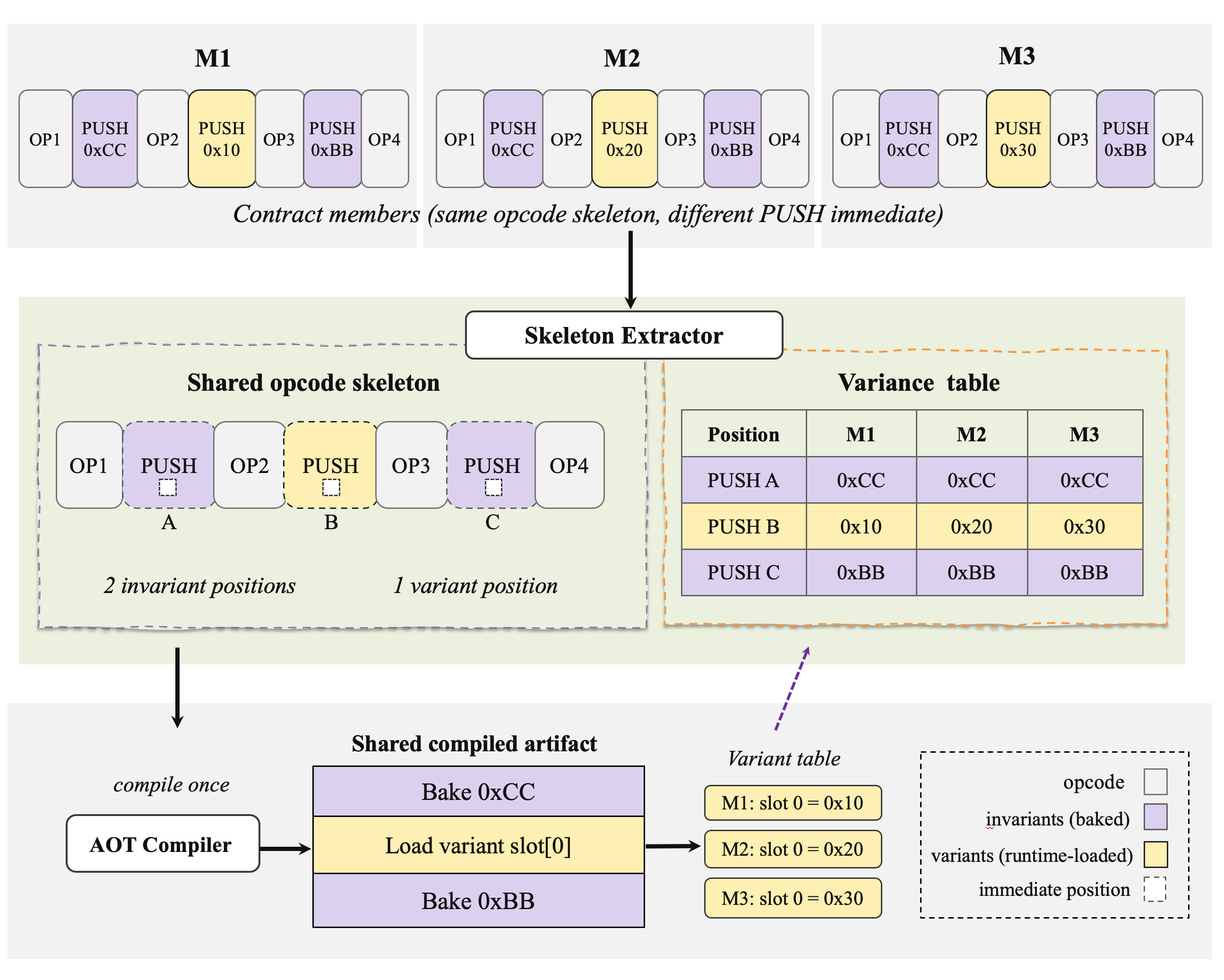}
  \caption{\textbf{\textsc{SkelOT} compilation pipeline.}}
\label{fig:design_overview}
\end{figure}

\textsc{SkelOT} raises the reuse unit of EVM AOT compilation from one deployed contract's code hash to the skeleton shared by a contract family. The design therefore shares one compiled artifact across family members while keeping invariant constants visible to the optimizer, preserving per-contract execution behavior, and leaving most of the host AOT pipeline unchanged. Figure~\ref{fig:design_overview} shows the \textsc{SkelOT} compilation pipeline applied to a contract family.

\subsection{Design Principles}
\label{sec:design:goals}

\textsc{SkelOT} is guided by one reuse objective. The objective is to maximize reuse of compiled structure across contracts in the same family, so that compilation effort for one member benefits every member sharing its skeleton. 

There are three design constraints. The first constraint is semantic preservation. Shared execution must match per-contract compilation in success or failure, gas, output, and post-state for every transaction. The second is runtime efficiency. Reuse should add no measurable overhead beyond noise. The third is deployability. The system should integrate with existing EVM AOT pipelines by reusing their standard bytecode translation, optimization, and native-code emission path rather than introducing a specialized execution engine.

\subsection{Choosing A Byte-Level Skeleton Representation}
\label{sec:design:skeleton}

Recall from \S\ref{sec:families:defn} that a skeleton is a contract's bytecode with PUSH immediates, the appended Solidity metadata section, and any trailing zero bytes removed, with every remaining opcode preserved in place; two contracts share a family when their skeleton byte sequences are bytewise identical. The representation is deliberately a lean byte sequence rather than a richer structural form such as an abstract syntax tree or a normalized control-flow graph. The host compiler's existing bytecode analysis already reconstructs the control-flow graph and instruction boundaries from the opcode sequence during translation, and sharing that reconstruction across family members is exactly the reuse \textsc{SkelOT} exploits. A richer skeleton would not expand what can be shared; it would only add transformation stages between the raw bytecode and the work the backend already performs.

\subsection{Skeleton Extraction And Family Membership}
\label{sec:design:extraction}

The skeleton extractor performs a single linear scan of the bytecode. Each opcode byte is emitted as-is, immediate bytes following a PUSH opcode are skipped, and the appended Solidity metadata section and any trailing zero bytes are removed, so deployments of the same logical contract that differ only in those trailing bytes share a single skeleton. Two contracts belong to the same family exactly when their skeleton byte sequences match. The prototype identifies each skeleton by a fast hash of its byte sequence~\cite{aumasson2012siphash}. The hash only locates candidates, and the prototype then confirms membership by comparing the skeleton bytes, so the family check stays exact at an $O(1)$ lookup cost, and the hash is the family identifier used by the caching and dispatch machinery of \S\ref{sec:implementation}. Byte equality also handles compiler-version skew. If two Solidity versions generate different code, the skeletons differ and the contracts never share an artifact.

The EVM draws no boundary between code and data, so stripping has to establish one. The trailer's last two bytes give its length. We read that length, check that the covered bytes have the CBOR shape of Solidity metadata, and only then strip them. A jump could still target a JUMPDEST byte inside the trailer, so we mark stripped positions as invalid jump targets in both the native code and our interpreter, and such a jump fails the same way on both sides. However, inside the code region there is no such boundary. If reachable code remains after a skeleton's last terminator, execution could run off the end into the stripped bytes, so we conservatively reject that skeleton (\S\ref{sec:results:rq2} reports the counts). EOF-style formats~\cite{beregszaszi2021eip3540} would separate code from data at the format level and remove the need for both checks.

Two consequences follow from this minimal definition. First, two contracts that both push the value zero but use different opcodes to do so (a dedicated zero-PUSH opcode versus a one-byte PUSH opcode with a zero immediate) end up in different families, because the skeleton preserves the opcode choice rather than normalizing across alternative encodings of the same value. This is the one point at which the design trades a little reuse for a locally decidable notion of family equality that needs no cross-contract normalization. Second, the constant classification below requires at least two members. A singleton family compiles identically to per-contract compilation, treating every PUSH immediate as a fixed compile-time constant, and gains no reuse benefit until a second member arrives and supplies the cross-member comparison that exposes variants.

\subsection{The Invariant/Variant Constant Split}
\label{sec:design:split}

We classify once, when a family forms, reading the bytecode of every member. A PUSH position is invariant if all members hold the same value there, and variant if any member differs. The family descriptor records this split. The compiler bakes the invariant values into one shared block of native code and leaves a table slot for each variant position. We number the slots in the order the variant positions appear in the skeleton, so every member fills a table with the same layout. We keep the descriptor fixed once written, so a later contract must pass admission (\S\ref{sec:design:admission}) to join the family; \S\ref{sec:discussion:admission} discusses more details. Algorithm~\ref{alg:implementation} lists the steps.

Our central design insight is \emph{invariant/variant constant split}. In real contract families, most PUSH immediates are identical across members; we call these positions \emph{invariants}. Only a small subset differs across members; we call these positions \emph{variants}. \textsc{SkelOT} handles them differently. Invariants remain compile-time constants in the generated code, preserving the backend's ability to fold constants, specialize surrounding code, and use them as static branch targets or comparison pivots. This folding is distinct from the folding that source-level compilers such as solc~\cite{solidity2026docs} perform while generating EVM bytecode, whose results are already frozen inside the PUSH immediates. The folding we mean happens one stage later, when the backend lowers bytecode to native code and can simplify around a literal immediate but not around a table load. \textsc{SkelOT} excludes variants from the compiled skeleton and loads them at runtime from a compact per-member table in a fixed skeleton-defined order. Thus, unlike per-contract compilation, which keeps every immediate constant but forgoes reuse, \textsc{SkelOT} retains the stable constants needed for optimization while externalizing only the values required for cross-member code sharing.

The opposite design point is an \emph{all-variant} scheme, which classifies every eligible immediate as a variant. This maximizes skeleton sharing but replaces compile-time constants with runtime table loads. We exclude immediates folded into control-flow targets, since changing them would invalidate the shared branch structure rather than simply update a table entry. For all remaining immediates, once the code loads a value from the per-member table, the backend can no longer fold arithmetic, specialize comparisons, or resolve branches around that value. These optimizations are a key reason why per-contract compilation outperforms interpretation on EVM workloads. An all-variant design therefore sacrifices compile-time specialization even for values that never vary across the family, leading to larger native artifacts, longer compile time, and extra runtime loads. We quantify these compile-side costs in \S\ref{sec:results:ablation}.

The observed structure of contract families motivates the split. Real families exhibit a stable skeleton with shallow variance, so preserving invariant positions as compile-time constants lets the backend keep the optimizations that matter for per-contract compilation. \S\ref{sec:results:rq1} quantifies this structure, showing that the multi-member population is overwhelmingly invariant, that the variant share shrinks further with contract size, and that the invariant classification is stable across the deployment horizon, not only within the measurement window. The ablation in \S\ref{sec:results:ablation} confirms that preserving invariants as compile-time constants accounts for much of the retained optimization benefit.

\subsection{Runtime Model}
\label{sec:design:runtime}

At runtime, every member of a family executes the same compiled artifact. Each member supplies separately a small per-member table holding its variant values, one entry per variant position. When execution reaches a variant position, the compiled code reads the matching entry from the current contract's table and pushes the 256-bit value, exactly as a plain PUSH of an invariant constant would. The deployed bytecode is never rewritten. We choose between an inline immediate and a table load only when lowering to native code, so offsets, jump targets, and gas stay untouched. The shared artifact therefore carries the skeleton, the invariant constants, and the optimizations built around them, while the per-member table carries only what the family allows to differ. Each variant PUSH instruction is compiled with a fixed slot index that locates its entry in the per-member table, and the same slot assignment applies to every member, so a single artifact serves all members without per-member code generation.

\subsection{Correctness Boundary}
\label{sec:design:correctness}

Let \(B_i\) be an admitted member of a family, let \(s=S(B_i)\) be its skeleton, and let \(x\) be a transaction input. For each PUSH position \(p\) in \(s\), \textsc{SkelOT} classifies \(p\) as invariant when every family member carries the same immediate at that position, and as variant otherwise. The descriptor assigns each variant position \(p\) a slot \(j(p)\) in appearance order, and member \(B_i\) supplies the table entry \(T_i[j(p)]\) containing exactly the immediate value at \(p\) in its original bytecode. The relative preservation claim is
\[
  \mathsf{Obs}(\mathsf{SkelOT}(s,T_i,B_i,x)) =
  \mathsf{Obs}(\mathsf{NaiveAOT}(B_i,x)),
\]
where \(\mathsf{Obs}\) records success or failure, gas, output, and post-state. The claim is relative. Assuming the underlying per-contract AOT compiler preserves the EVM semantics formalized by prior work~\cite{cassez2023dafny}, \textsc{SkelOT} preserves the behavior of that compiler while changing the reuse unit.

\emph{Proof (sketch).} The proof sketch is a case analysis over the executed instruction stream. Skeleton equality fixes the opcode layout, instruction boundaries, and PUSH widths for every family member. For any non-PUSH instruction, \textsc{SkelOT} and Naive AOT therefore execute the same EVM opcode at the same logical position. For an invariant PUSH position, \textsc{SkelOT} embeds the same 256-bit value that Naive AOT would have embedded for \(B_i\). For a variant PUSH position \(p\), \textsc{SkelOT} loads \(T_i[j(p)]\), defined to be the same 256-bit value the original bytecode of \(B_i\) would have pushed. Thus each semantic step sees the same opcode and the same stack operands in both executions. An induction over the trace gives identical observable behavior, provided the host compiler does not bake a value into native control flow and later allow that value to vary.
\qed

\smallskip
\noindent\textbf{Gas preservation.} Gas equality follows from the same step-by-step argument. At step \(k\), \textsc{SkelOT} and Naive AOT execute the same EVM opcode with the same stack operands. Static gas is therefore the same, since it is charged by opcode in the original instruction stream. Dynamic gas is also the same, since memory growth, copying, storage, and calls are all computed from the same operands in the same EVM context. The variant-table read does not change this argument. It happens inside the LLVM artifact before the EVM PUSH value reaches the stack, so it is not itself an EVM instruction and creates no separate gas event. Once the value reaches the stack, it is exactly the original immediate from \(B_i\), so any later gas formula that depends on that value receives the same input under \textsc{SkelOT} and Naive AOT.

\smallskip
\noindent\textbf{Code environment.} Code observation can also be stated in terms of the environment each opcode sees. The native artifact is only the execution vehicle. The EVM context presented to the compiled code still contains \(B_i\) as the current-frame bytecode, along with the same memory, gas meter, return-data buffer, and host state that Naive AOT uses. For current-frame code queries, \texttt{CODESIZE} and \texttt{CODECOPY} read the executing member's bytecode length and bytes from the per-call EVM context rather than from any compile-time constant, so the same artifact returns the correct length and bytes for every family member regardless of metadata or trailing-zero differences. Queries about another account, including \texttt{EXTCODESIZE}, \texttt{EXTCODECOPY}, and \texttt{EXTCODEHASH}, go through the same host-state lookup as Naive AOT and observe the code, length, or hash stored for the addressed account. \textsc{SkelOT} therefore changes the native artifact selected for execution, but not the EVM-visible code environment.

\smallskip
\noindent\textbf{Folded control flow.} The final boundary concerns constants that the compiler has already turned into a native branch structure. If a pushed value is recognized as a static jump target, the compiler may bake that target directly into the native branch. That value can no longer vary safely across family members. A later member could supply a different jump target in its bytecode, but the shared native code would still branch to the baked target from the earlier member. \textsc{SkelOT} therefore keeps such positions outside the variant set. If a family member needs one of those values to differ, that member lies outside the current shared-artifact boundary rather than being admitted with an unsound table entry.

Concretely, if a pushed jump target is folded into a native branch for one family member, reusing that branch for another member whose target differs would jump to the first member's target. \textsc{SkelOT} fences that PUSH position from the variant set, so the family either shares an artifact with the target fixed or does not share an artifact at all.

\subsection{Admission}
\label{sec:design:admission}

The admission mechanism determines whether a newly observed contract can join an existing family and avoid a full recompile. Given the contract bytecode and the family's invariant/variant classification, the admission descriptor checks two conditions: the contract skeleton must match the family skeleton, and every invariant position must contain the family's expected value. If both checks pass, the contract is admitted, and its per-member table is filled with the constants from its variant positions.

Admission is intentionally conservative. If a candidate disagrees with an invariant position, \textsc{SkelOT} rejects that candidate from the existing family rather than rewriting the descriptor, changing the variant set, or migrating the compiled artifact. This keeps the online rule aligned with the batch correctness boundary. Versioned families and online reclassification could recover more reuse, but they require an artifact invalidation policy, which we leave to \S\ref{sec:discussion:admission}. An alternative \emph{all-variant} configuration of \textsc{SkelOT}, evaluated in \S\ref{sec:results:ablation} and discussed in \S\ref{sec:discussion:admission}, trades a larger artifact and longer compile time for a more permissive admission rule over contracts that share the skeleton and do not require folded control-flow operands to vary.

\section{Implementation}
\label{sec:implementation}

\noindent\textbf{Compiler integration.} The \textsc{SkelOT} prototype adds family-level reuse to an existing EVM ahead-of-time compiler at four points. It extracts a skeleton key, classifies invariant and variant PUSH positions, compiles one shared artifact per reusable skeleton, and lowers only variant PUSH positions to indexed table loads. The host compiler's bytecode analysis, control-flow reconstruction, optimization passes, and native object emission remain unchanged.

\begin{algorithm}[t]
\caption{Preparing shared artifacts and member tables}
\label{alg:implementation}
\footnotesize
\begin{algorithmic}[1]
\Require Contracts $C$ and host ahead-of-time compiler $H$
\Ensure Family registry $R$ and member map $M$
\State $R \gets \emptyset$; $M \gets \emptyset$
\State $\mathcal{G} \gets$ the partition of $C$ by skeleton hash.
\ForAll{family group $G \in \mathcal{G}$}
  \If{$|G| = 1$}
    \State Let $c$ be the only contract in $G$.
    \State $a_c \gets H.\mathsf{compileContract}(c)$
    \State $M[\mathsf{codeHash}(c)] \gets \mathsf{Direct}(a_c)$ \Comment{\textcolor{violet}{singleton path}}
    \State \textbf{continue}
  \EndIf
  \State $s \gets \mathsf{skeletonHash}(G)$
  \State $D \gets \mathsf{classifyPushes}(G)$ \Comment{\textcolor{violet}{split invariant and variant PUSHes}}
  \State Choose a representative contract $r \in G$.
  \State $a_s \gets H.\mathsf{compileFamily}(r,D)$ \Comment{\textcolor{violet}{attempt shared compilation}}
  \If{$a_s = \bot$}
    \ForAll{contract $c \in G$}
      \State $a_c \gets H.\mathsf{compileContract}(c)$
      \State $M[\mathsf{codeHash}(c)] \gets \mathsf{Direct}(a_c)$
    \EndFor
    \State \textbf{continue} \Comment{\textcolor{violet}{fall back to direct artifacts}}
  \EndIf
  \State $R[s] \gets (a_s,D)$ \Comment{\textcolor{violet}{register the shared artifact}}
  \ForAll{contract $c \in G$}
    \State $T_c \gets \emptyset$
    \ForAll{$(p,j) \in \mathsf{Variants}(D)$}
      \State $T_c[j] \gets \mathsf{pushImmediate}(c,p)$
    \EndFor
    \State $M[\mathsf{codeHash}(c)] \gets \mathsf{Shared}(s,T_c)$ \Comment{\textcolor{violet}{member table}}
  \EndFor
\EndFor
\State \Return $(R, M)$
\end{algorithmic}
\end{algorithm}

\smallskip
\noindent\textbf{Artifact and member state.} Compiled artifacts and member data live in two tiers. A family registry, keyed by the skeleton hash, stores each successfully shared native object with its descriptor. A member map, keyed by contract code hash, stores either a direct per-contract artifact or a shared-family key with the member's variant table. Dispatch looks up the map entry. For a shared entry, it passes the member's table pointer in the call context. Each call carries its own context, so nested calls into the same family each read their own table. The shared artifact stays stateless. Executing a family member needs no per-member code generation and no per-PUSH dispatch. The shared artifact carries the skeleton, the invariant constants, and the optimizations built around them, while the per-member table carries only what the family allows to differ.

\smallskip
\noindent\textbf{Preparation path.} Algorithm~\ref{alg:implementation} summarizes this preparation path. The procedure groups contracts by skeleton hash and treats single-member groups as ordinary per-contract compilation (lines~1--9). For a reusable group, it builds a family descriptor that records the invariant PUSH immediates and the table slots for variant PUSH positions (lines~10--13). If the host compiler cannot build a shared artifact for that descriptor, the group falls back to per-contract artifacts (lines~14--20).

When shared compilation succeeds, the registry stores the shared artifact and its descriptor once under the skeleton key (line~21). The member map then stores one entry per contract (lines~22--27). Each entry points to the shared skeleton and carries that contract's concrete variant table, so dispatch can reuse the shared code while supplying the member-specific PUSH values.

\textsc{SkelOT} populates the registry only when shared compilation succeeds, and populates the member map for every compiled contract. Direct entries point to ordinary per-contract artifacts. Shared entries point through the family key and carry the concrete table used by that member.

\section{Evaluation}
\label{sec:eval-setup}
\label{sec:results}

The evaluation addresses four research questions. 

\begin{itemize}[leftmargin=*]
    \item \textbf{RQ1:} How common are reusable contract families in real EVM workloads?
    \item \textbf{RQ2:} How much does \textsc{SkelOT} reduce compilation cost and artifact size, and where do the savings come from?
    \item \textbf{RQ3:} Does \textsc{SkelOT} preserve correctness and maintain competitive dispatch time?
    \item \textbf{RQ4:} How much native code is needed to cover different shares of execution time?
\end{itemize}
 
Our primary baseline is Naive (per-hash) AOT, implemented in the same LLVM-based AOT pipeline as \textsc{SkelOT}. Both build on revmc~\cite{paradigm2024revmc}, and the baseline is revmc's own per-hash compilation path, run at the same compiler version, optimization level, and thread count. This choice isolates the design variable, the unit at which compiled code is reused. Direct wall-clock comparisons with other EVM execution engines would mix this variable with implementation language, intermediate representation design, compiler backend, cache organization, and JIT or AOT policy. We treat external engines such as evmone and DTVM as contextual systems in \S\ref{sec:related:engines} rather than as causal baselines.

We use Base mainnet as the primary corpus because it is a high-volume EVM-compatible chain with an active contract-deployment ecosystem, which stresses both compile-side and runtime-side metrics under realistic load. We arbitrarily sample a window of 10{,}000 blocks, blocks 38{,}004{,}930 through 38{,}014{,}929 inclusive, containing 3.52M transactions. Three blocks are empty, so the experiments materialize 9{,}997 block state pre-images. EOF-format bytecode is deferred in the EVM roadmap, so we exclude empty bytecode and bytecode beginning with \texttt{0xEF}, leaving 19{,}130 unique compilable bytecodes. To verify that this arbitrary choice is representative, we check the window in two ways. Skeleton reduction rates stay within narrow ranges across ten 1K-block sub-windows on every chain (\S\ref{sec:families}), and the top families span months to years of deployment (Table~\ref{tab:table3_deployment_spans_xchain}), so the window captures persistent workload structure.

Experiments run on an Intel Xeon Platinum 8275CL with sixteen hardware threads and 30~GiB DRAM under Ubuntu 24.04. Both modes are built with rustc 1.91 and LLVM 21.1 and run with sixteen worker threads. Object footprint is the cumulative byte length of the LLVM-emitted \texttt{.o} files per mode. The runtime then links and loads these objects as \texttt{.so} files; we measure the \texttt{.o} stage, identically for both modes. We report per-member runtime variant tables separately as runtime metadata. The variant-table payload totals 5.4 MiB across 9{,}910 members of multi-member families, with a median table size of 352 bytes per member. The full registry file additionally stores a per-member hash identifier and per-family bookkeeping fields, totaling 9.6 MiB on disk.

\begin{figure}[!]
  \centering
  \includegraphics[width=\columnwidth]{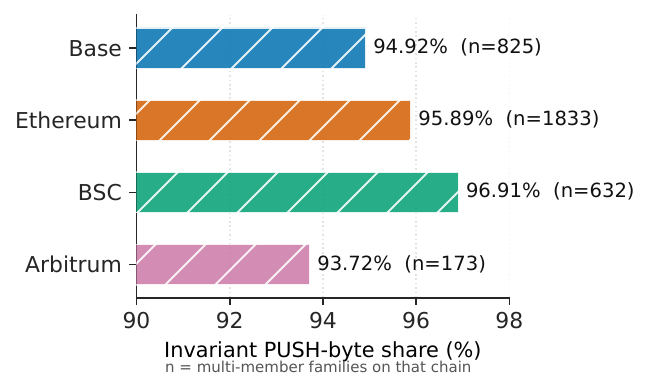}
  \input{figures/fig4_push_invariance_xchain_caption.tex}
\end{figure}

\subsection{RQ1: Workload Structure}
\label{sec:results:rq1}

RQ1 asks whether the invariant/variant split reflects a recurring workload property rather than a Base-specific artifact. Across all four studied chains, multi-member families have the same qualitative shape. Only a small fraction of PUSH-immediate bytes vary across family members. The byte-weighted variant share is 5.08\% on Base, 4.11\% on Ethereum, 3.09\% on BSC, and 6.28\% on Arbitrum, computed over 825, 1{,}833, 632, and 173 multi-member families respectively. Correspondingly, 93.72\%--96.91\% of embedded-constant bytes are invariant and can be baked into the shared skeleton as compile-time constants. The split is decided per position, and 99.52\% of PUSH positions on Base are invariant, so the byte-weighted share above is a conservative figure. Larger contracts also vary less. In an average family from the tiny bucket, 12.6\% of the PUSH bytes are variant. In the huge bucket, the average is 0.4\%. Figure~\ref{fig:push_invariance_xchain} reports the four-chain breakdown. These results show that the invariant/variant split matches the common case in deployed contracts. Most constants stay the same across members, so the per-contract runtime table carries only a small residual.

\begin{table}[!]
  \centering
  \footnotesize
  \caption{Deployment-time spans for top skeleton families across four chains.}
  \label{tab:table3_deployment_spans_xchain}
  \begin{tabular}{rcc}
    \toprule
    \multicolumn{1}{c}{\textbf{Family}} & \#\textbf{members} & \textbf{Span} (days) \\
    \cmidrule(lr){1-3}
    Clanker ERC-20 [Base] & 465 & 99.4 \\
    EIP-1167 proxy [ARB] & 33 & 1,690.1 \\
    EIP-1167 proxy [Base] & 342 & 805.8 \\
    EIP-1167 proxy [BSC] & 160 & 1,912.7 \\
    EIP-1167 proxy [ETH] & 629 & 1,732.7 \\
    ERC-1155 creator [ETH] & 167 & 887.4 \\
    ERC-20 template [BSC] & 164 & 57.8 \\
    PancakeSwap LmPool [ARB] & 28 & 835.4 \\
    PancakeSwap LmPool [Base] & 212 & 763.7 \\
    PancakeSwap LmPool [BSC] & 243 & 973.2 \\
    PancakeSwap V3 [ARB] & 37 & 971.4 \\
    PancakeSwap V3 [Base] & 402 & 789.0 \\
    PancakeSwap V3 [BSC] & 1{,}123 & 1,090.9 \\
    Swap helper [ETH] & 127 & 670.2 \\
    Transparent proxy [ARB] & 45 & 267.1 \\
    Transparent proxy [ETH] & 164 & 534.5 \\
    Uniswap V3 Pool [ARB] & 116 & 1,762.5 \\
    Uniswap V3 Pool [Base] & 3{,}323 & 752.9 \\
    Uniswap V3 Pool [BSC] & 405 & 1,142.3 \\
    Uniswap V3 Pool [ETH] & 2{,}722 & 1,355.1 \\
    \bottomrule
  \end{tabular}
\end{table}

The same family structure is durable over deployment time, so it is not merely a short-window co-occurrence effect. Table~\ref{tab:table3_deployment_spans_xchain} reports the five largest multi-member families on each chain and traces sampled members through full chain history using archive RPC. The top Base families span 99--806 days, Ethereum reaches 1{,}733 days, BSC reaches 1{,}913 days, and Arbitrum reaches 1{,}762 days. The long-span families include Uniswap V3 and PancakeSwap V3 pool skeletons, EIP-1167 proxies, transparent proxies, and token or collection templates. This temporal evidence supports the interpretation that the invariant-heavy families observed in the corpora are persistent deployment templates, not contracts that merely appear together in short measurement windows. The same spans also support the classification. Members deployed years apart hold the same values at every invariant position, so the invariant sets have stayed stable.

\subsection{RQ2: Compile-Side Savings And Attribution}
\label{sec:results:rq2}

For the compile-side benchmark, we compile the filtered corpus once under Naive AOT and once under \textsc{SkelOT} from an empty object cache. We measure compile units, wall-clock compile time, object footprint, and fallback count.

\begin{figure}[b]
  \centering
  \includegraphics[width=\columnwidth]{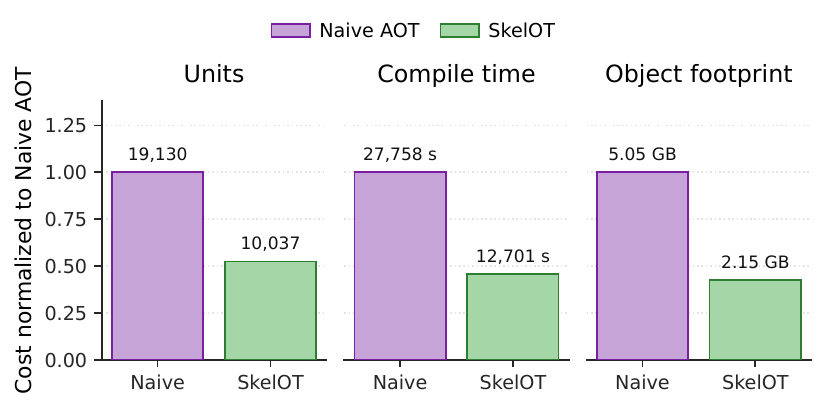}
  \input{figures/fig5_compile_cost_caption.tex}
\end{figure}

\textsc{SkelOT} reduces (Figure~\ref{fig:compile_cost}) the number of compile units by 47.5\%, shrinks the native-object footprint by 57.4\%, and cuts compile time by 2.19$\times$. Fallback is small. A conservative check on where the stripped code region ends rejects 7 skeletons covering 17 members, and one family of 3 members degrades because its members jump to different static targets, so these 20 contracts compile per-hash instead.

To separate skeleton sharing from the invariant/variant split, we run a three-mode compile-side ablation over all 817 shared families. \label{sec:results:ablation} \emph{Naive AOT} compiles each member on its own, and we sum the per-member cost from the full per-hash build. \emph{SkelOT} shares one skeleton artifact and keeps invariant immediates as compile-time constants. \emph{All-variant} is \textsc{SkelOT} with the classification step skipped, so it still shares the skeleton but routes every eligible PUSH immediate through the per-contract table.

\begin{figure}[tbp]
  \centering
  \includegraphics[width=\columnwidth]{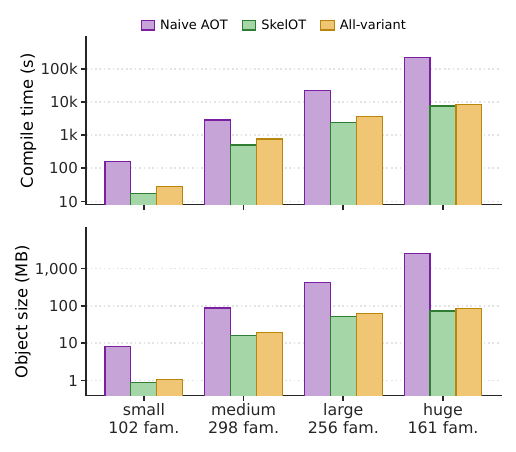}
  \input{figures/fig6_ablation_caption.tex}
\end{figure}

Figure~\ref{fig:ablation} shows both parts of the benefit. The Naive bars measure family sharing relative to \textsc{SkelOT}, with compile-time ratios from $5.7\times$ to $29.6\times$ and object-size ratios from $5.4\times$ to $34.7\times$ across buckets. Compiling each family once takes 10{,}385~s instead of 248{,}239~s and produces 141.4~MB instead of 3{,}040~MB, a $23.9\times$ and $21.5\times$ difference over the 9{,}910 members. The all-variant bars hold skeleton sharing fixed and isolate the constant-classification policy, with compile-time ratios from $1.11\times$ to $1.58\times$ and object-size ratios up to $1.21\times$. The invariant/variant split contributes a measurable compile-side benefit beyond skeleton sharing alone.

Both gains grow with contract size. The huge bucket alone holds about half of the members. Naive AOT would spend most of its 3{,}040~MB on this bucket, while \textsc{SkelOT} serves it with 73~MB. Larger contracts have more instructions per skeleton, so skipping one recompilation saves more work. They also hold more invariant constants, which keeps more values folded into the shared code.

The all-variant regression has a concrete per-PUSH mechanism. Families with more PUSH positions forced to be variants produce larger native artifacts. Removing compile-time constants creates extra native code, rather than only reducing the optimizer's ability to simplify existing code.

On Arbitrum, our least favorable corpus, the same effect holds at smaller scale, cutting compilation units by 23.1\%, footprint by 26.2\%, and compile time by $1.29\times$.

New contracts keep arriving after the initial corpus is compiled. Over the later windows, every 1{,}000 blocks bring 764 new unique bytecodes on average. Under per-hash reuse, every one of them is a fresh compilation candidate. Under \textsc{SkelOT}, at least 44.6\% join an existing family and reuse its artifact at once, so the candidate pool shrinks by nearly half.

\subsection{RQ3: Correctness And Runtime Behavior}
\label{sec:results:rq3}

Because paired timing is expensive, we evaluate runtime behavior by replaying every hundredth block of the corpus under both Naive AOT and \textsc{SkelOT} dispatch. Across 99 sampled blocks, we record paired per-member timings for 1{,}842 invoked family members after 2 warmup and 5 measured rounds per mode. For each member we run a paired $t$-test on the per-round differences and correct for testing all 1{,}842 hypotheses at once with the Benjamini-Hochberg procedure at $q = 0.05$~\cite{benjamini1995controlling}. For each measured family member, speedup is the mean Naive AOT dispatch time divided by the mean \textsc{SkelOT} dispatch time over the measured rounds, so the metric compares per-member dispatch time during block replay.

To confirm execution-level correctness, we replay every non-empty block of the corpus once under Naive AOT and once under \textsc{SkelOT} dispatch and compare each transaction's success/failure status, gas consumption, and output between the two modes. Across all 9{,}997 non-empty blocks (3.52M transactions), \textsc{SkelOT} dispatch produces byte-identical execution results to Naive AOT, with zero mismatches. The replay also covers the hardest case for table binding, where a call into one family member runs inside a call into another member of the same family and the two frames must read different tables. The corpus contains 3.4M such nested family frames, and all replay with zero mismatches.

Runtime timing shows no aggregate regression against Naive AOT. In the paired per-contract comparison, \textsc{SkelOT} has a 1.31$\times$ median per-member speedup, with a short slowdown tail. The median dispatch time falls from 23.08~$\mu$s under Naive AOT to 18.12~$\mu$s. After the Benjamini-Hochberg correction, 1{,}171 contracts are faster, 644 equivalent, and 27 slower, with a bootstrap 95\% confidence interval on the median speedup of [1.286, 1.345]. Figure~\ref{fig:runtime_speedup} shows the full per-pair distribution. Most of the distribution lies near or above the no-change boundary, with larger gains in the right tail, and every sampled block has a block-internal median above 1.0$\times$. We attribute this 1.31$\times$ median speedup to instruction-fetch locality under cross-instance code-region reuse. The counter evidence in \S\ref{sec:discussion:implications} supports this instruction-cache explanation.

\begin{figure}[!]
  \centering
  \includegraphics[width=\columnwidth]{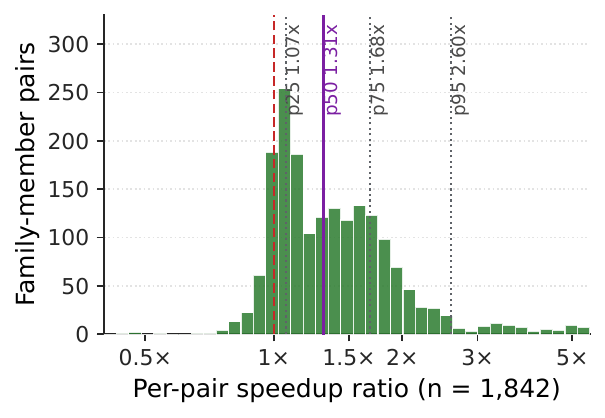}
  \input{figures/fig7_runtime_speedup_distribution_caption.tex}
\end{figure}

Table~\ref{tab:e2e_replay} gives the broader picture, the wall-clock cost of replaying all 10{,}000 blocks under each mode, including compilation.

\begin{table}[!]
  \centering
  \scriptsize
  \setlength{\tabcolsep}{4pt}
  \begin{threeparttable}
  \caption{End-to-end replay of all 10K blocks, three modes.}
  \label{tab:e2e_replay}
  \begin{tabular}{lccccc}
    \toprule
    \textbf{Mode} & \textbf{Exec (s)} & \textbf{Family (s)} & \textbf{Other (s)} & \textbf{Compile (s)} & \textbf{Fam.\ build (s)} \\
    \midrule
    Interpreter & 534.4 & --- & --- & --- & --- \\
    Naive AOT   & 459.1 & 93.8 & 365.2 & 27{,}758.3 & --- \\
    SkelOT      & 450.6 & 84.8 & 365.8 & 12{,}701.5 & 1.1 \\
    \bottomrule
  \end{tabular}
  \end{threeparttable}
\end{table}

Compilation accounts for most of the difference between Naive AOT and \textsc{SkelOT}. Execution time also favors \textsc{SkelOT}, and the gain is isolated. Family-member frames drop from 93.8~s to 84.8~s, while state access, transaction assembly, and precompiled-contract execution stay flat, since compilation strategy touches none of them. Family-table construction adds 1.1~s, negligible next to either compile cost.

\subsection{RQ4: Budget Efficiency}
\label{sec:results:rq4}

Budget efficiency asks how much native code an AOT system must materialize and store to cover a target share of execution time. For each target, we take the smallest execution-weighted \textsc{SkelOT} skeleton prefix that reaches the target. That prefix defines both the shared artifacts \textsc{SkelOT} compiles and the concrete code hashes those artifacts serve. To compare against Naive AOT on the same deployment scope, we count how many native artifacts Naive AOT would need under its one-artifact-per-code-hash policy, and compare the cumulative artifact footprint. We also report a more conservative reading, in which Naive AOT ranks contracts by their own execution time and compiles only its own hot set.

Figure~\ref{fig:coverage_budget} shows that \textsc{SkelOT} serves the same deployment scope with substantially fewer native artifacts and lower footprint across the measured coverage range. The hottest part of the workload shows the largest gap. At 75\% execution-time coverage, 35 \textsc{SkelOT} artifacts cover 3{,}933 distinct code hashes, while Naive AOT needs one artifact per hash. This is a $112\times$ reduction in artifact count and a $437\times$ reduction in footprint. The advantage remains large deeper in the ranking, with count and footprint compression of $33.9\times$ and $168\times$ at 90\% coverage, and $6.2\times$ and $62\times$ at 99\%. Under the conservative reading, Naive AOT needs $1.9\times$ the artifacts and $5.9\times$ the footprint at 75\%, and $2.5\times$ and $13.9\times$ at 95\%.

Compression decreases as the coverage target moves into the long tail, where more singleton families enter the prefix and each additional \textsc{SkelOT} artifact serves fewer additional deployments. Even there, footprint compression stays above artifact-count compression, which indicates that the hot shareable families near the head are also expensive for Naive AOT to compile separately. \textsc{SkelOT} thus spends compilation budget where it matters first, on highly executed templates that would otherwise produce many large per-hash artifacts. The 99\% point in Figure~\ref{fig:coverage_budget} is slightly below full coverage of the compile corpus because a small set of compile units did not execute in the measured trace. The conservative reading instead widens, from $1.6\times$ at 50\% to $2.5\times$ at 99\% in artifacts, because contracts without a family dominate the head of the workload while families sit in the mid and long tail.

\begin{figure}[tbp]
  \centering
  \includegraphics[width=\linewidth]{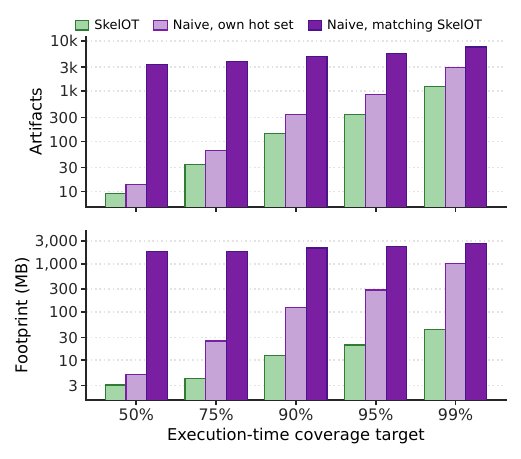}
  \input{figures/fig8_coverage_budget_caption.tex}
\end{figure}

\section{Discussion}
\label{sec:discussion}

Family-level reuse changes the operating point of EVM AOT systems, not only their compile budget. This section interprets the measured gains, the workloads that support them, and the boundary that still limits transfer.

\subsection{Implications For EVM AOT Systems}
\label{sec:discussion:implications}

A code hash remains the right way to name a deployed contract, but it is too fine-grained as the sole unit of AOT compilation. On the Base corpus, compiling at family granularity cuts compilation units, native-artifact footprint, and compile time while preserving per-contract dispatch identity. The budget result sharpens the same point. A finite AOT budget should buy coverage over deployed behavior, not redundant native objects for members of the same template family. Runtime measurements on the same corpus are also favorable, with no measured penalty and a 1.31$\times$ median speedup over Naive AOT.

When Naive AOT compiles each contract on its own, a block that runs many similar contracts keeps jumping to different native binaries. Each jump can make the CPU bring in another piece of code before execution can continue. Those waits are the main source of the runtime gap. \textsc{SkelOT} instead gives all members of the same family one shared binary. Once that binary is hot, later calls to other family members reuse the same code in the instruction cache.

The hardware counters match this explanation. We use the CPU's top-level slot accounting because its four buckets add up cleanly. Almost the entire reduction comes from slots where the CPU is waiting for code to arrive at the front of the pipeline. The other buckets do not explain the speedup. Backend waiting increases slightly, wasted speculative work falls a little, and useful retired work is nearly unchanged (Table~\ref{tab:runtime_attribution_tma}). More detailed counters point to the same cause. L1 instruction-cache stalls fall by about 1.0K and 0.25K cycles per call, and L2 code-read misses also fall. Counters for frontend decode bandwidth move by only a few cycles, and off-core code-read cycles are zero. Our machine does not expose the CPU events needed to split the frontend bucket cleanly into smaller additive pieces, so we use these detailed counters only as supporting evidence for an instruction-cache explanation. We read counters per call over the same fixed block order in both modes, repeat each measurement several times after warmup, and report averaged paired deltas in Table~\ref{tab:runtime_attribution_tma}. We do not pin CPU frequency or ASLR; to remove that noise, we use repeated paired measurements.

The same mechanism predicts where \textsc{SkelOT} will not help, and the data agrees. A shared binary only stays in cache when it is hit often enough; a family with only a handful of members generates too few hits to keep its code warm. \textsc{SkelOT} also pays a small fixed cost per call. It loads variant PUSH constants from a per-member data table rather than inlining them as immediate operands. When the family is large, this cost disappears against the cache savings; when the family is small, the cost is real and the savings never materialize. In our measurements on the Base corpus, the runtime ratio rises sharply with family size. Families with fewer than one hundred members stay essentially flat near 1.07$\times$, and only families of at least one hundred members produce the 1.58$\times$ median that drives the headline. The contracts on which \textsc{SkelOT} actually slows things down sit almost entirely in the small-family stratum, exactly where the mechanism predicts no benefit.

A practical AOT system should therefore enable family compilation only when a family is large enough to carry its weight. Below that threshold, plain per-contract compilation is the safer choice; above it, the locality win compounds with every additional member. The exact cutoff depends on the calling profile and on how much per-call work the host EVM does outside the JIT'd code, but our bucket-level evidence puts a sensible default in the range of roughly one hundred members. With this admission rule in place, \textsc{SkelOT} keeps its wins on the long-tail high-reuse families that drive the headline number and avoids the regressions on small families.

\begin{table}[tbp]
  \centering
  \scriptsize
  \setlength{\tabcolsep}{4pt}
  \caption{Top-level TMA slot attribution for SkelOT versus Naive AOT. Values are SkelOT minus Naive; negative values are savings.}
  \label{tab:runtime_attribution_tma}
  \begin{tabular}{lrr}
    \toprule
    \textbf{TMA bucket} & \textbf{Delta} & \textbf{Interpretation} \\
    & \textbf{(slots/call)} & \\
    \midrule
    Frontend bound & $-8{,}149.7$ & fewer frontend undersupply slots \\
    Backend bound  & $+165.0$     & offset: more backend-bound slots \\
  Bad speculation & $-141.4$    & fewer wasted speculative slots \\
  Retiring       & $-10.3$      & nearly unchanged useful-work slots \\
  \midrule
  \textbf{Total slots} & $\mathbf{-8{,}136.5}$ & net slot reduction \\
  \bottomrule
\end{tabular}
\end{table}

\subsection{When \textsc{SkelOT} Helps}
\label{sec:discussion:when}

\textsc{SkelOT} helps most when deployments reuse the same bytecode structure and differ only in a small set of immediates. That is the common case in our measurements, especially for large contracts. When a workload has less repeated structure, \textsc{SkelOT} exposes less sharing and falls back toward fine-grained compilation. The current boundary is intentionally precise. Differences in structure, control flow, or opcode choice separate families. Future abstractions can admit more cases while keeping the same rule. Compiled code should be shared only when member-specific behavior is explicitly represented.

Similarly, \textsc{SkelOT} does not normalize PUSH widths. Such normalization can raise reuse but is not semantics-free in the EVM. PUSH width affects instruction boundaries and jump targets. Metadata and byte length are observable through code-observation instructions, so \textsc{SkelOT} strips metadata and trailing zero bytes only while preserving the original code view at execution. A safe normalization layer for opcode encodings would need the same separation between the reuse key and the EVM-visible bytecode. \textsc{SkelOT} leaves that larger design point to follow-on work and keeps the present reuse predicate byte-level and auditable.

Families benefit in different ways. In some, a few hot members~\cite{adams2021uniswap,pancakeswap2023contracts} carry most of the execution time. Later members with the same structure and different constants get native code at once, with no new compilation. In others, execution time is spread over many members~\cite{wang2026factories,diangelo2019mayflies,diangelo2020wallets,khan2022clones}. One compilation and one artifact then serve the whole family. In every case, the members differ only in constructor parameters that deployment writes into the code.

\subsection{External Validity}
\label{sec:discussion:external}

The evidence has two scopes. Workload-structure measurements use one 10K-block window from each of Base, Ethereum, BSC, and Arbitrum, and show the same invariant-heavy family structure across chains. Compile-side savings, all-block correctness replay, timing, and hardware-counter attribution use Base as the primary corpus on a revmc-based LLVM AOT prototype under a fixed version of the EVM rules. A smaller Arbitrum compile run confirms the compile-side direction (\S\ref{sec:results:rq2}). Within the Base window, family structure is stable across sub-windows and over deployment time (\S\ref{sec:results:rq1}). Prior studies of factory deployment, proxy contracts, composition, on-chain dependencies, and bytecode clones report that family-structured deployment extends well beyond this window~\cite{wang2026factories,sun2023composition,zhang2025proxy,jin2025onchain,he2020clones}. Broader replication should quantify how compile-side ratios, runtime attribution, and budget compression transfer across fork rules, protocol mixes, chains, and compiler backends. The design requirement is narrower. A backend must keep invariant PUSH values as compile-time constants, lower variant PUSH values as table loads, and bind the current member's table during dispatch.

\begin{table*}[t]
  \centering
  \footnotesize
  \setlength{\tabcolsep}{2.5pt}
  \renewcommand{\arraystretch}{1.0}
  \newcommand{\capnote}[2]{#1\,{\scriptsize(#2)}}
  \begin{threeparttable}
    \caption{Summary and comparison of related works.}
    \label{tab:table5_related_positioning}
    \begin{tabular}{r p{2.3cm} p{2.3cm}p{2.3cm}p{2.3cm} C{3.5cm}}
      \toprule
      \multicolumn{1}{c}{\multirow{2}{*}{\textbf{Works}}} & \multicolumn{1}{c}{\multirow{2}{*}{\textbf{Refs.}}} & \multicolumn{3}{c}{\textbf{Capability supplied}} & \multirow{2}{*}{\textbf{Role} for SkelOT} \\
      \cmidrule(lr){3-5}
      & & \multicolumn{1}{c}{\textbf{Native exec.}} & \multicolumn{1}{c}{\textbf{Family signal}} &  \multicolumn{1}{c}{\textbf{Reuse}} & \\
      \midrule
      AOT/native engines
        & \cite{paradigm2024revmc,ipsilon2024evmone,zhou2025dtvm}
        & \capnote{\protect\symP}{AOT/native}
        & \capnote{\protect\symN}{per hash}
        & \capnote{\protect\symS}{artifact reuse}
        & execution substrate \\
      Workload studies
        & \cite{wang2026factories,sun2023composition,zhang2025proxy,he2020clones,diangelo2024evolution}
        & \protect\symN{}
        & \capnote{\protect\symP}{factories/proxies}
        & \capnote{\protect\symN}{no compiler}
        & opportunity evidence \\
      Clone detection
        & \cite{liu2019birthmarks,liu2018eclone,gao2019smartembed,wang2025clonedetection,mo2025solidity}
        & \capnote{\protect\symN}{orthogonal}
        & \capnote{\protect\symS}{clusters}
        & \capnote{\protect\symN}{not units}
        & boundary warning \\
      Reusable compilation
        & \cite{futamura1999partial,fallin2025partial,mehta2023jitreuse,pecimuth2024reusability,pecimuth2025ir}
        & \capnote{\protect\symN}{not EVM AOT}
        & \capnote{\protect\symN}{no families}
        & \capnote{\protect\symP}{context reuse}
        & conceptual ancestry \\
      Below-translation reuse
        & \cite{bruening2003adaptive,bellard2005qemu,reddi2007persistent,tallam2010safeicf,lld2024icf}
        & \capnote{\protect\symS}{native blocks}
        & \capnote{\protect\symN}{post-translation}
        & \capnote{\protect\symP}{dedup/cache}
        & contrast point \\
      \midrule
      \multicolumn{1}{c}{\textbf{SkelOT}}
        & \multicolumn{1}{c}{-}
        & \capnote{\protect\symP}{EVM AOT}
        & \capnote{\protect\symP}{skeleton}
        & \capnote{\protect\symP}{family reuse}
        & family-aware AOT reuse\\
      \bottomrule
    \end{tabular}
    \begin{tablenotes}[flushleft]
      \scriptsize
      \item \quad Legend: \protect\symP{} supplied; \protect\symS{} partial; \protect\symN{} not supplied.
    \end{tablenotes}
  \end{threeparttable}
  \vspace{-5pt}
\end{table*}

\subsection{Admission Policy Trade-Offs}
\label{sec:discussion:admission}

We run classification and table construction in a batch phase, which matches how AOT works. Family classification only adds a comparison over the compile batch. We checked whether our way of choosing families is stable over time. We split the window in half, and families chosen from the first half alone cover 80.2\% of execution time in the second half, while the gold-standard answer is 80.8\%. The classification also goes stale slowly, since the deployment spans in \S\ref{sec:results:rq1} show top families staying on chain for months to years.

This paper evaluates batch family reuse, but the same design choice also appears at deployment time. A deployed system needs an online rule for newly arriving contracts: either preserve the compiled artifact exactly and admit only members that match its descriptor, or widen the descriptor so that more same-skeleton contracts can join. \textsc{SkelOT}'s current rule is deliberately conservative. A new contract joins an existing compiled family only when its skeleton matches and every invariant position preserves the family's expected value. When a candidate changes a current invariant, \textsc{SkelOT} rejects that contract from the family rather than rewriting the shared artifact online. This boundary keeps admission monotone, keeps the cache key stable after compilation, and avoids artifact invalidation or migration. The cost is that some same-skeleton contracts remain outside the already compiled family.

A more permissive endpoint is unconditional admission through an all-variant classification. Under this rule, every eligible PUSH value is lowered as a runtime table load, so any contract with the same skeleton can join an existing family by binding its own table at deployment; only diverging static-jump targets still block admission (\S\ref{sec:design:correctness}). Figure~\ref{fig:admission_tradeoff} sketches the difference in a listing-style example aligned with Figure~\ref{fig:skelot-Intro}. Moving an otherwise stable immediate from the native artifact into the table removes a compile-time constant, which can prevent constant folding, comparison specialization, and other backend simplifications around that value. Over the shared family artifacts, where the two designs differ, all-variant takes 22.2\% longer to compile, produces 19.4\% larger artifacts, runs family member frames 18.2\% slower, and grows the registry file from 9.6~MiB to 526~MiB. The current split therefore favors deployments that can amortize a stricter admission boundary across many same-skeleton members, whereas unconditional admission favors deployments that prioritize join-on-arrival simplicity over per-family efficiency.

\begin{figure}[t]
  \centering
  \begingroup
  \definecolor{skelotPanel}{RGB}{247,247,247}
  \definecolor{skelotFrame}{RGB}{221,221,221}
  \definecolor{skelotInnerFrame}{RGB}{188,188,188}
  \definecolor{skelotHi}{RGB}{255,239,179}

  \newcommand{\varval}[1]{%
    \begingroup
    \setlength{\fboxsep}{1pt}%
    \colorbox{skelotHi}{\strut #1}%
    \endgroup}

  \newcommand{\rejectmark}{\textcolor{symred}{\ding{55}}}
  \newcommand{\admitmark}{\textcolor{symgreen}{\ding{51}}}

  \newcommand{\innerbox}[1]{%
    \begingroup
    \setlength{\fboxsep}{2pt}%
    \setlength{\fboxrule}{0.25pt}%
    \noindent\hfill
    \fcolorbox{skelotInnerFrame}{skelotPanel}{%
      \begin{minipage}{0.92\linewidth}
        #1
      \end{minipage}}%
    \hfill\null
    \par
    \endgroup}

  \newcommand{\codepanel}[2]{%
    \begingroup
    \setlength{\fboxsep}{3pt}%
    \setlength{\fboxrule}{0.25pt}%
    \fcolorbox{skelotFrame}{skelotPanel}{%
      \begin{minipage}[t]{0.405\columnwidth}
        {\footnotesize\bfseries #1}\par
        \vspace{2pt}
        {\scriptsize\ttfamily\linespread{0.92}\selectfont\raggedright
        #2}
      \end{minipage}}%
    \endgroup}

  \codepanel{Current admission}{%
    \innerbox{%
    shared artifact:\par
    \quad token0 := load table[0]\par
    \quad token1 := load table[1]\par
    \quad fee := 500 baked\par
    }%
    \vspace{3pt}
    \innerbox{%
    contract A table:\par
    \quad slot0=\varval{0x10}; slot1=\varval{0xaa}\par
    \quad slot2=\varval{500} \admitmark\par
    }%
    \vspace{3pt}
    \innerbox{%
    contract C table:\par
    \quad slot0=\varval{0x30}; slot1=\varval{0xcc}\par
    \quad slot2=\varval{3000} \rejectmark\par
    }%
  }%
  \hfill
  \codepanel{All-variant admission}{%
    \innerbox{%
    shared artifact:\par
    \quad token0 := load table[0]\par
    \quad token1 := load table[1]\par
    \quad fee := load table[2]\par
    }%
    \vspace{3pt}
    \innerbox{%
    contract A table:\par
    \quad slot0=\varval{0x10}; slot1=\varval{0xaa}\par
    \quad slot2=\varval{500} \admitmark\par
    }%
    \vspace{3pt}
    \innerbox{%
    contract C table:\par
    \quad slot0=\varval{0x30}; slot1=\varval{0xcc}\par
    \quad slot2=\varval{3000} \admitmark\par
    }%
  }%

  \vspace{-2pt}
  \caption{\textbf{Admission trade-off}. Split rejects a candidate changing a baked invariant (\textit{left}); all-variant admission accepts it by moving the position into the member table (\textit{right}).}
  \Description{Two listing-style panels compare admission rules. In the left panel, token0 and token1 come from a table while fee is baked as 500. Contract A stores 0x10, 0xaa, and 500, while contract C stores 0x30, 0xcc, and 3000, marked with a red cross because the changed fee is rejected. In the right panel, fee also comes from the table, so the same contract C values are marked with a green check and admitted.}
  \label{fig:admission_tradeoff}
  \endgroup
\end{figure}

Two possible improvements look promising. One is to recompute families periodically, regrouping the contracts rejected since the last pass. This needs a policy for retiring live artifacts. The other is to declare which constants vary at the source level. The classification is then known when the first member arrives.

Deduplicating the artifact store is another promising optimization. Content-defined chunking~\cite{muthitacharoen2001lbfs,xia2016fastcdc} splits binaries at boundaries derived from the content, so similar binaries share most chunks on disk. Our prototype does not use it, but it could shrink the on-disk store further.

Versioned families are a possible middle ground beyond the current design. They could recover some currently rejected members without pushing every position into the runtime table, for example by compiling a second artifact when a once-invariant position begins to vary. That extra flexibility would need a policy for when to fork a family, how to route old and new members, and whether to retire or migrate existing artifacts. An arrival-order replay could compare conservative admission, unconditional admission, and versioned families in terms of acceptance rate, artifact churn, compile cost, and runtime effect.

\section{Related Work}
\label{sec:related}

We summarize \textsc{SkelOT} and relevant studies in Table~\ref{tab:table5_related_positioning}.

\noindent\textbf{EVM execution and AOT compilation.}
\label{sec:related:engines} Ahead-of-time compilation of EVM bytecode to native code is an established
engineering direction. The \texttt{revmc} project lowers EVM bytecode to
LLVM IR and produces one native artifact per code
hash~\cite{paradigm2024revmc}; \texttt{evmone} offers an optional AOT
path~\cite{ipsilon2024evmone}; and DTVM explores a hybrid lazy-JIT
architecture on a Wasm/dMIR substrate~\cite{zhou2025dtvm}. \textsc{SkelOT} inherits the LLVM-based AOT execution model but changes the compilation-reuse unit from the full bytecode or code hash to a contract-family skeleton, as
detailed in \S\ref{sec:design}. DTVM's design axis is orthogonal to the reuse granularity we study.

\vspace{3pt}
\noindent\textbf{Empirical structure of EVM workloads.}
\label{sec:related:workloads}
Production EVM contracts exhibit substantial structural reuse. Factory-driven deployment accounts for more than 90\,\% of contracts created since 2020~\cite{wang2026factories}; subcontracts often cluster around recurring imported patterns~\cite{sun2023composition}; and code-sharing proxies and clones are widely used~\cite{zhang2025proxy,he2020clones}. These studies show that many deployed contracts come from recurring templates, motivating family-level optimization, but they stop at workload characterization rather than compilation reuse.

DiAngelo et al.~\cite{diangelo2024evolution} are closest to our abstraction. They use bytecode skeletons to avoid repeated large-scale weakness analysis of bytecodes that differ only outside the chosen abstraction. \textsc{SkelOT} builds on the same observation but changes the boundary's role. For AOT compilation, skeleton equality alone is insufficient. The compiler must decide which constants can remain baked into shared native code and which must be externalized into per-contract tables. \textsc{SkelOT} turns skeletons from a static-analysis boundary into an AOT reuse boundary through an invariant/variant constant split and an EVM-specific correctness condition.

\vspace{3pt}
\noindent\textbf{Smart-contract clone detection.}
\label{sec:related:clones} Ethereum clone detection is mature but orthogonal to compilation. The representative three approaches (i.e., birthmark-based, semantic sketch, structural embedding) identify similar contracts or clone clusters, but they do not define executable reuse units~\cite{liu2019birthmarks,liu2018eclone,gao2019smartembed,wang2025clonedetection,mo2025solidity}. For AOT reuse under EVM gas equivalence, similarity is not enough. The system must know which constants can be shared safely and which must remain contract-specific. \textsc{SkelOT} supplies this missing correctness layer through its invariant/variant constant split.

\vspace{3pt}
\noindent\textbf{Reusable compilation and JIT-code reuse.}
\label{sec:related:reuse}
Reusable compilation is conceptually related to \textsc{SkelOT}, but it usually assumes a different correctness model. Partial evaluation specializes a program with respect to known inputs~\cite{futamura1999partial,fallin2025partial}. Recent JIT-reuse systems reduce repeated compilation by reusing compiled functions, optimized IR, or specialization results across runs and virtual-machine instances~\cite{mehta2023jitreuse,pecimuth2024reusability,pecimuth2025ir}.

These systems target dynamic runtimes, where reused code can be guarded, replay-validated, or deoptimized when the context changes. \textsc{SkelOT} works in a stricter EVM AOT setting. Shared native code must preserve per-contract behavior and gas accounting before execution. We therefore place reuse at the contract-family skeleton and share code only through an explicit invariant/variant split.
\section{Conclusion}
\label{sec:conclusion}


We show that code hash is too fine-grained as the reuse unit for EVM AOT compilation. Modern EVM workloads contain contract families that share instruction skeletons while differing only in a few embedded constants. We design \textsc{SkelOT} to exploit this structure by compiling one artifact per family, baking invariant constants into shared code, and loading only variants from per-contract tables. 

Across four EVM-compatible chains, we find that 23.1--47.6\% of unique compilable bytecodes collapse into shared skeletons. On Base, \textsc{SkelOT} cuts compilation units by 47.5\%, artifact footprint by 57.4\%, and compile time by $2.19\times$, while preserving Naive-AOT behavior and achieving a $1.31\times$ median runtime speedup. 


\bibliographystyle{ACM-Reference-Format}
\bibliography{refs}

@inproceedings{liu2019birthmarks,
  author    = {Liu, Han and Yang, Zhiqiang and Jiang, Yu and Zhao, Wenqi and Sun, Jiaguang},
  title     = {Enabling Clone Detection For {Ethereum} Via Smart Contract Birthmarks},
  booktitle = {IEEE/ACM International Conference on Program Comprehension (ICPC)},
  pages     = {105--115}, 
  year      = {2019}, 
}

@inproceedings{he2020clones,
  author    = {He, Ningyu and Wu, Lei and Wang, Haoyu and Guo, Yao and Jiang, Xuxian},
  title     = {Characterizing Code Clones in the {Ethereum} Smart Contract Ecosystem},
  booktitle = {Financial Cryptography and Data Security (FC)}, 
  pages     = {654--675}, 
  year      = {2020}, 
}

@article{khan2022clones,
  author    = {Khan, Faizan and David, Istvan and Varro, Daniel and McIntosh, Shane},
  title     = {Code Cloning in Smart Contracts on the {E}thereum Platform: An Extended Replication Study},
  journal   = {IEEE Transactions on Software Engineering (TSE)},
  volume    = {49},
  number    = {4},
  pages     = {2006--2019},
  year      = {2022}, 
}

@article{wang2026factories,
  author  = {Wang, Ziyue and Shen, Zongwen and Chen, Lei and Song, Wei and Ge, Jidong and Huang, LiGuo and Luo, Bin},
  title   = {Empirical Analysis of Smart Contract Factories on {EVM}-compatible Chains},
  journal = {ACM Transactions on Software Engineering and Methodology (TOSEM)},
  pages   = {3787216},
  year    = {2026},
}

@inproceedings{diangelo2019mayflies,
  author    = {di Angelo, Monika and Salzer, Gernot},
  title     = {Mayflies, Breeders, and Busy Bees in {E}thereum: Smart Contracts Over Time},
  booktitle = {Proceedings of the 3rd ACM Workshop on Blockchains, Cryptocurrencies and Contracts (BCC)},
  pages     = {1--10},
  year      = {2019},
}

@inproceedings{diangelo2020wallets,
  author    = {di Angelo, Monika and Salzer, Gernot},
  title     = {Characteristics of Wallet Contracts on {E}thereum},
  booktitle = {2nd Conference on Blockchain Research \& Applications for Innovative Networks and Services (BRAINS)},
  pages     = {232--239},
  year      = {2020},
}

@article{diangelo2024evolution,
  author  = {di Angelo, Monika and Durieux, Thomas and Ferreira, Jo{\~a}o F. and Salzer, Gernot},
  title   = {Evolution of Automated Weakness Detection in {Ethereum} Bytecode: A Comprehensive Study},
  journal = {Empirical Software Engineering (ESE)},
  volume  = {29},
  number  = {2},
  pages   = {41},
  year    = {2024}, 
}

@misc{paradigm2024revmc,
  author       = {{Paradigm}},
  title        = {{revmc}: {JIT} and {AOT} Compiler for the {Ethereum} Virtual Machine},
  year         = {2024},
  howpublished = {\url{https://github.com/paradigmxyz/revmc}},
}

@misc{ipsilon2024evmone,
  author       = {{Ipsilon}},
  title        = {{evmone}: Fast {Ethereum} Virtual Machine Implementation},
  year         = {2024},
  howpublished = {\url{https://github.com/ipsilon/evmone}},
  note         = {Apache-2.0 licensed open-source software; successor to the Ewasm/{ethereum}/evmone repository; accessed 2026-04-22}
}

@misc{zhou2025dtvm,
  author       = {Zhou, Wei and Xu, Xiong and Wei, Changzheng and Yan, Ying and Tang, Wei and Chen, Zhihao and Huang, Xuebing and others},
  title        = {{DTVM}: Revolutionizing Smart Contract Execution with Determinism and Compatibility},
  year         = {2025},
  eprint       = {2504.16552},
  archivePrefix = {arXiv},
  primaryClass = {cs.DC},
  note         = {\url{http://arxiv.org/abs/2504.16552v2}}
}

@inproceedings{sun2023composition,
  author    = {Sun, Kairan and Xu, Zhengzi and Liu, Chengwei and Li, Kaixuan and Liu, Yang},
  title     = {Demystifying the Composition and Code Reuse in {Solidity} Smart Contracts},
  booktitle = {Proceedings of the {ACM} Joint European Software Engineering Conference and Symposium on the Foundations of Software Engineering (ESEC/FSE)},
  year      = {2023}, 
}

@inproceedings{zhang2025proxy,
  author    = {Zhang, Mengya and Shukla, Preksha and Zhang, Wuqi and Zhang, Zhuo and Agrawal, Pranav and Lin, Zhiqiang and Zhang, Xiangyu and Zhang, Xiaokuan},
  title     = {An Empirical Study of Proxy Contracts at the {Ethereum} Ecosystem Scale},
  booktitle = {Proceedings of the IEEE/ACM International Conference on Software Engineering (ICSE)},
  year      = {2025}, 
}

@inproceedings{liu2018eclone,
  author    = {Liu, Han and Yang, Zhiqiang and Liu, Chao and Jiang, Yu and Zhao, Wenqi and Sun, Jiaguang},
  title     = {{EClone}: Detect Semantic Clones in {Ethereum} via Symbolic Transaction Sketch},
  booktitle = {Proceedings of the {ACM} Joint Meeting on European Software Engineering Conference and Symposium on the Foundations of Software Engineering (ESEC/FSE) --- Demo/Tool Track},
  year      = {2018}, 
}

@inproceedings{gao2019smartembed,
  author    = {Gao, Zhipeng and Jayasundara, Vinoj and Jiang, Lingxiao and Xia, Xin and Lo, David and Grundy, John},
  title     = {{SmartEmbed}: A Tool for Clone and Bug Detection in Smart Contracts through Structural Code Embedding},
  booktitle = {IEEE International Conference on Software Maintenance and Evolution (ICSME)},
  year      = {2019}, 
}

@article{wang2025clonedetection,
  author  = {Wang, Zuobin and Wan, Zhiyuan and Chen, Yujing and Zhang, Yun and Lo, David and Xie, Difan and Yang, Xiaohu},
  title   = {Clone Detection for Smart Contracts: How Far Are We?},
  journal = {Proceedings of the ACM on Software Engineering (ASE/FSE)},
  volume  = {2}, 
  pages   = {1249--1269},
  year    = {2025}, 
}

@article{futamura1999partial,
  author  = {Futamura, Yoshihiko},
  title   = {Partial Evaluation of Computation Process---An Approach to a Compiler-Compiler},
  journal = {Higher-Order and Symbolic Computation},
  volume  = {12},
  number  = {4},
  pages   = {381--391},
  year    = {1999}, 
  note    = {Reprint of the original 1971 paper in Systems, Computers, Controls 2(5):45--50.}
}

@inproceedings{fallin2025partial,
  author    = {Fallin, Chris and Bernstein, Maxwell},
  title     = {Partial Evaluation, Whole-Program Compilation},
  booktitle = {Proceedings of the {ACM SIGPLAN} Conference on Programming Language Design and Implementation (PLDI)},
  year      = {2025}, 
}

@article{mehta2023jitreuse,
  author  = {Mehta, Meetesh Kalpesh and Kry{\'n}ski, Sebasti{\'a}n and Gualandi, Hugo Musso and Thakur, Manas and Vitek, Jan},
  title   = {Reusing Just-in-Time Compiled Code},
  journal = {Proceedings of the {ACM} on Programming Languages (PACMPL)},
  volume  = {7},
  number  = {OOPSLA2},
  pages   = {1176--1197},
  year    = {2023},
}

@inproceedings{pecimuth2024reusability,
  author    = {Pe{\v c}im{\'u}th, Andrej and Leopoldseder, David and T{\r u}ma, Petr},
  title     = {An Analysis of Compiled Code Reusability in Dynamic Compilation},
  booktitle = {Proceedings of the {ACM SIGPLAN} International Workshop on Virtual Machines and Intermediate Languages (VMIL)},
  year      = {2024}, 
}

@inproceedings{pecimuth2025ir,
  author    = {Pe{\v c}im{\'u}th, Andrej and Leopoldseder, David and T{\r u}ma, Petr},
  title     = {Reusing Highly Optimized {IR} in Dynamic Compilation},
  booktitle = {Proceedings of the 39th European Conference on Object-Oriented Programming (ECOOP)},
  year      = {2025}, 
}

@inproceedings{cassez2023dafny,
  title={Formal and executable semantics of the ethereum virtual machine in dafny},
  author={Cassez, Franck and Fuller, Joanne and Ghale, Milad K and Pearce, David J and Quiles, Horacio MA},
  booktitle={International Symposium on Formal Methods (FM)},
  pages={571--583},
  year={2023},
  organization={Springer}
}

@misc{jin2025onchain,
  author        = {Jin, Xiangfu and Liu, Zihao and Monperrus, Martin},
  title         = {On-Chain Analysis of Smart Contract Dependency Risks on {E}thereum},
  year          = {2025},
  eprint        = {2503.19548},
  archivePrefix = {arXiv},
  primaryClass  = {cs.SE},
  note          = {arXiv:2503.19548}
}

@inproceedings{mo2025solidity,
  author    = {Mo, Ran and Song, Haopeng and Ding, Wei and Wu, Chaochao},
  title     = {Code Cloning in {S}olidity Smart Contracts: Prevalence, Evolution, and Impact on Development},
  booktitle = {Proceedings of the {IEEE/ACM} International Conference on Software Engineering ({ICSE})},
  pages     = {3060--3071}, 
  year      = {2025}, 
}

@article{jantz2013jitpolicy,
  author  = {Jantz, Michael R. and Kulkarni, Prasad A.},
  title   = {Exploring Single and Multilevel {JIT} Compilation Policy for Modern Machines},
  journal = {ACM Transactions on Architecture and Code Optimization (TACO)},
  volume  = {10},
  number  = {4},
  pages   = {40:1--40:29},
  year    = {2013},
}

@misc{wasmtime_cache_docs,
  author       = {{Bytecode Alliance}},
  title        = {Cache Configuration of \texttt{wasmtime}},
  howpublished = {\url{https://docs.wasmtime.dev/cli-cache.html}},
  year         = {2026},
  note         = {Accessed 2026-04-24}
}

@misc{v8_codecache_devs,
  author       = {Swirski, Leszek},
  title        = {Code Caching for {JavaScript} Developers},
  howpublished = {\url{https://v8.dev/blog/code-caching-for-devs}},
  year         = {2019},
  note         = {Accessed 2026-04-24}
}

@inproceedings{aumasson2012siphash,
  title={SipHash: a fast short-input PRF},
  author={Aumasson, Jean-Philippe and Bernstein, Daniel J},
  booktitle={International Conference on Cryptology in India (INDOCRYPT)},
  pages={489--508},
  year={2012},
  organization={Springer}
}

@article{benjamini1995controlling,
  author  = {Benjamini, Yoav and Hochberg, Yosef},
  title   = {Controlling the False Discovery Rate: A Practical and Powerful Approach to Multiple Testing},
  journal = {Journal of the Royal Statistical Society: Series B (Methodological)},
  volume  = {57},
  number  = {1},
  pages   = {289--300},
  year    = {1995},
}

@misc{wood2014ethereum,
  author       = {Wood, Gavin},
  title        = {Ethereum: A Secure Decentralised Generalised Transaction Ledger},
  year         = {2014},
  howpublished = {Ethereum Project Yellow Paper},
  note         = {\url{https://ethereum.github.io/yellowpaper/paper.pdf}; Accessed 2026-04-28}
}

@inproceedings{lattner2004llvm,
  author    = {Lattner, Chris and Adve, Vikram},
  title     = {{LLVM}: A Compilation Framework for Lifelong Program Analysis \& Transformation},
  booktitle = {Proceedings of the International Symposium on Code Generation and Optimization ({CGO})},
  pages     = {75--86},  
  year      = {2004}, 
}

@article{cytron1991ssa,
  author    = {Cytron, Ron and Ferrante, Jeanne and Rosen, Barry K. and Wegman, Mark N. and Zadeck, F. Kenneth},
  title     = {Efficiently Computing Static Single Assignment Form and the Control Dependence Graph},
  journal   = {ACM Transactions on Programming Languages and Systems (TOPLAS)},
  volume    = {13},
  number    = {4},
  pages     = {451--490},
  year      = {1991}, 
}

@misc{coinbase2023base,
  author       = {{Coinbase}},
  title        = {Base: An {Ethereum} {L2} Built on the {OP} {S}tack},
  year         = {2023},
  howpublished = {\url{https://docs.base.org/}},
  note         = {Mainnet launched 2023-08-09; Accessed 2026-04-28}
}

@misc{bnbchain2024overview,
  author       = {{BNB Chain}},
  title        = {{BNB} Smart Chain: High Performance {DeFi} Hub},
  year         = {2026},
  howpublished = {{BNB Chain Documentation}},
  note         = {\url{https://docs.bnbchain.org/bnb-smart-chain/overview/}; Accessed 2026-05-06}
}

@inproceedings{li2025does,
  title={Does finality gadget finalize your block? A case study of {B}inance consensus},
  author={Li, Rujia and Ding, Jingyuan and Wang, Qin and Jia, Keting and Zhang, Haibin and Duan, Sisi},
  booktitle={34th USENIX Security Symposium (USENIX Sec)},
  pages={4109--4125},
  year={2025}
}

@misc{offchainlabs2024arbitrum,
  author       = {Bousfield, Lee and Bousfield, Rachel and Buckland, Chris and Burgess, Ben and Colvin, Joshua and Felten, Edward W. and Goldfeder, Steven and Goldman, Daniel and Huddleston, Braden and Kalodner, Harry and Lacs, Frederico Arnaud and Ng, Harry and Sanghi, Aman and Wilson, Tristan and Yermakova, Valeria and Zidenberg, Tsahi},
  title        = {{Arbitrum Nitro}: A Second-Generation Optimistic Rollup},
  year         = {2022},
  howpublished = {{Arbitrum Documentation}},
  note         = {\url{https://docs.arbitrum.io/nitro-whitepaper.pdf}; Accessed 2026-05-06}
}

@misc{adams2021uniswap,
  author       = {Adams, Hayden and Zinsmeister, Noah and Salem, Moody and Keefer, River and Robinson, Dan},
  title        = {{Uniswap} v3 {Core}},
  year         = {2021},
  howpublished = {Whitepaper},
  note         = {\url{https://app.uniswap.org/whitepaper-v3.pdf}; Accessed 2026-04-28}
}

@misc{vogelsteller2015erc20,
  author       = {Vogelsteller, Fabian and Buterin, Vitalik},
  title        = {{EIP-20}: {Token Standard}},
  year         = {2015},
  howpublished = {{Ethereum Improvement Proposal}},
  note         = {\url{https://eips.ethereum.org/EIPS/eip-20}; Accessed 2026-04-28}
}

@misc{murray2018eip1167,
  author       = {Murray, Peter and Welch, Nate},
  title        = {{EIP-1167}: {Minimal Proxy Contract}},
  year         = {2018},
  howpublished = {{Ethereum Improvement Proposal}},
  note         = {\url{https://eips.ethereum.org/EIPS/eip-1167}; Accessed 2026-04-28}
}

@misc{pancakeswap2023contracts,
  author       = {{PancakeSwap}},
  title        = {{PancakeSwap} {V3} Contracts},
  year         = {2023},
  howpublished = {{GitHub} repository},
  note         = {\url{https://github.com/pancakeswap/pancake-v3-contracts}; Accessed 2026-05-06}
}

@inproceedings{muthitacharoen2001lbfs,
  author    = {Muthitacharoen, Athicha and Chen, Benjie and Mazi{\`e}res, David},
  title     = {A Low-Bandwidth Network File System},
  booktitle = {Proceedings of the {ACM} Symposium on Operating Systems Principles ({SOSP})},
  year      = {2001}
}

@inproceedings{xia2016fastcdc,
  author    = {Xia, Wen and Zhou, Yukun and Jiang, Hong and Feng, Dan and Hua, Yu and Hu, Yuchong and Liu, Qing and Zhang, Yucheng},
  title     = {{FastCDC}: A Fast and Efficient Content-Defined Chunking Approach for Data Deduplication},
  booktitle = {Proceedings of the {USENIX} Annual Technical Conference ({ATC})},
  year      = {2016}
}

@misc{solidity2026docs,
  author       = {{Solidity Team}},
  title        = {{Solidity} Documentation},
  year         = {2026},
  howpublished = {Online documentation},
  note         = {\url{https://docs.soliditylang.org}; Accessed 2026-08-24}
}

@misc{openzeppelin2024proxy,
  author       = {{OpenZeppelin}},
  title        = {Proxy},
  year         = {2026},
  howpublished = {{OpenZeppelin} Contracts Documentation},
  note         = {\url{https://docs.openzeppelin.com/contracts/5.x/api/proxy}; Accessed 2026-05-06}
}

@misc{clanker2024clankertoken,
  author       = {{Clanker}},
  title        = {{ClankerToken} v3.1.0 and v4.0.0},
  year         = {2025},
  howpublished = {{Clanker} Documentation},
  note         = {\url{https://clanker.gitbook.io/clanker-documentation/references/core-contracts/clankertoken-v3.1.0-and-v4.0.0}; Accessed 2026-05-06}
}

@misc{radomski2018eip1155,
  author       = {Radomski, Witek and Cooke, Andrew and Castonguay, Philippe and Therien, James and Binet, Eric and Evans, Ron},
  title        = {{EIP-1155}: Multi Token Standard},
  year         = {2018},
  howpublished = {{Ethereum Improvement Proposal}},
  note         = {\url{https://eips.ethereum.org/EIPS/eip-1155}; Accessed 2026-05-06}
}

@misc{manifold2024creator,
  author       = {{Manifold}},
  title        = {{Manifold Creator}},
  year         = {2025},
  howpublished = {{Manifold} Documentation},
  note         = {\url{https://docs.manifold.xyz/v/manifold-for-developers/manifold-creator-architecture/overview}; Accessed 2026-05-06}
}

@inproceedings{bruening2003adaptive,
  author    = {Bruening, Derek and Garnett, Timothy and Amarasinghe, Saman},
  title     = {An Infrastructure for Adaptive Dynamic Optimization},
  booktitle = {Proceedings of the International Symposium on Code Generation and Optimization ({CGO})},
  pages     = {265--275}, 
  year      = {2003}, 
}

@inproceedings{bellard2005qemu,
  author       = {Bellard, Fabrice},
  title        = {{QEMU}, a Fast and Portable Dynamic Translator},
  booktitle    = {Proceedings of the {USENIX} Annual Technical Conference (ATC), {FREENIX} Track},
  pages        = {41--46},
  publisher    = {USENIX Association}, 
  year         = {2005}, 
}

@inproceedings{reddi2007persistent,
  author    = {Reddi, Vijay Janapa and Connors, Dan and Cohn, Robert and Smith, Michael D.},
  title     = {Persistent Code Caching: Exploiting Code Reuse Across Executions and Applications},
  booktitle = {Proceedings of the International Symposium on Code Generation and Optimization ({CGO})},
  pages     = {74--88}, 
  year      = {2007}, 
}

@article{xie2026mhot,
  title={MHOT: Height-Optimized Authenticated Data Structure for Blockchain State Commitment},
  author={Xie, Sipeng and Wu, Qianhong and Li, Minghang and Gao, Qiyuan and Qin, Bo and Wang, Qin},
  journal={arXiv preprint arXiv:2606.11736; USENIX Security 2026, to appear},
  year={2026}
}

@misc{tallam2010safeicf,
  author       = {Tallam, Sriraman and Coutant, Cary and Taylor, Ian Lance and Li, Xinliang David and Demetriou, Chris},
  title        = {Safe {ICF}: Pointer Safe and Unwinding Aware Identical Code Folding in the {Gold} Linker},
  howpublished = {GCC Developers' Summit 2010},
  year         = {2010},
  note         = {\url{https://research.google/pubs/safe-icf-pointer-safe-and-unwinding-aware-identical-code-folding-in-gold/}; Accessed 2026-04-28}
}

@misc{lld2024icf,
  author       = {{LLVM Project}},
  title        = {{lld}: The {LLVM} Linker},
  year         = {2026},
  howpublished = {\url{https://lld.llvm.org/}},
  note         = {Linker documentation including identical code folding ({ICF}); Accessed 2026-04-28}
}

@misc{beregszaszi2021eip3540,
  author       = {Beregszaszi, Alex and Bylica, Pawe{\l} and Maiboroda, Andrei},
  title        = {{EIP-3540}: {EOF} - {EVM} Object Format v1},
  year         = {2021},
  howpublished = {{Ethereum Improvement Proposal}},
  note         = {\url{https://eips.ethereum.org/EIPS/eip-3540}; Accessed 2026-09-01}
}

\end{document}